\documentclass[aps,pra,reprint,groupedaddress]{revtex4-2}

\usepackage{graphicx}
\usepackage{dcolumn} 
\usepackage{bm}      

\begin{document}

\title{
Quantum correlations via vector soliton interactions
}


\author{Andrey Konyukhov}
\email[]{KonukhovAI@sgu.ru}
\affiliation{Institute of Physics, Saratov State University, 
Astrakhanskaya 83, 410012 Saratov, Russia}


\date{\today}

\begin{abstract}
The generation of quantum-correlated pulse pairs in a dispersion modulated birefringent fiber is considered. The photon-number correlations and squeezing are studied using linearized quantum fluctuation theory. Two models of the pulse propagation in an optical fiber are used. The first model is based on the Manakov equations, and the second one is based on the coupled nonlinear Schr\"odinger equations with differential group delay and birefringence terms. In the Manakov model the correlated pulse pairs can be generated using splitting of a second-order soliton and inelastic collision of two fundamental solitons. The interpulse correlations depends on the modulation period of the fiber dispersion. In the model of the coupled nonlinear Schr\"odinger equations the correlated pulse pair can be produced using pulse splitting due to polarization mode dispersion. The pulses have orthogonal polarization states. This allows the pulses to be separated into two different channels using a polarization beam splitter. The interpulse correlation depends on the interplay between polarization mode dispersion and polarization instability.  
\end{abstract}


\maketitle

\section{Introduction}

Continuous-variable optical quantum entanglement and correlations play an essential role in various quantum applications, such as quantum communications \cite{Zhang2021,Zhang2024,ZZhang2024}, quantum computation \cite{Menicucci2006, Asavanant2024}, and quantum imaging \cite{Gregory2021, Massar2023}. Entangled quantum systems are correlated when each system is measured. A particular type of entanglement can be created by interfering two squeezed beams on a beam splitter \cite{Kim2002}. 
Carter \textit{et al.} proposed a way to generate the squeezed light by using optical solitons \cite{Carter1987}. Optical solitons exhibit quadrature fluctuations below the level of vacuum fluctuations. Fluctuations are reduced over the soliton bandwidth.
The observation of multimode quantum correlations in fiber optical solitons was reported in \cite{Spalter1998, Leuchs2002}. Kerr nonlinearity couples fluctuations at different frequencies and provides photon-number correlations between different spectral components of the pulse. The measured frequency-domain correlation matrix exhibit a characteristic butterfly pattern reflecting the internal quantum noise structure of a fundamental soliton. 

The bound higher-order solitons governed by the nonlinear Schr\"odinger (NLS) equation show substantial reduction in fluctuations of soliton masses, phases, and positions compared to the fundamental soliton \cite{Marchukov2020}. The enhancement of narrow-bandwidth spectral quantum correlations in high-order solitons was considered in \cite{Schmidt2000, Lee2005a}. The correlation matrices show domains with almost perfect positive and negative correlations, which produce a rectangular pattern with boundaries at frequencies of about $\pm T_0^{-1}$, where $T_0$ is the width of the fundamental soliton.

Entangled states can be generated by colliding Schr\"odinger solitons. Two co-propagating in-phase solitons attract and collide \cite{Agrawal2013}. After collision, the solitons become quantum entangled in the sense that their quadrature components satisfy the inseparability criterion \cite{Lai2009}. The collision of two solitons traveling at different group velocities due to different carrier frequencies exhibits transient interpulse photon number correlations \cite{Konig2002}. Strong interpulse quantum correlations appear as the solitons approach each other. The interpulse correlations disappear after elastic collision of solitons. Permanent interpulse correlation can be achieved via cross-phase modulation \cite{Hirosawa2007}. A strong negative correlation in photon number arises between a soliton and a trapped pulse, which propagate at different carrier frequency. 

The concept of polarization entanglement for continuous variables was introduced in \cite{Korolkova2002}. A light beam formed by the interference of two orthogonally polarized quadrature-squeezed beams exhibits squeezing in some of the Stokes parameters. In \cite{Korolkova2002}, the scheme was implemented using a fiber optic Sagnac interferometer. 

Quantum entangled and squeezed light can be generated directly by propagating a vector soliton in an optical fiber \cite{Corney2008, Sorokin2021}. At low input pulse energy, the squeezing ratio is limited by the depolarizing effect of Brillouin scattering. At high pulse energy, the squeezing ratio is reduced by stimulated Raman scattering. Variations in the zero-dispersion wavelength and nonlinear polarization mode dispersion produce excess noise \cite{Chapman2023}. Optical pulses with an average position accuracy beyond the standard quantum limit can be generated by adiabatically expanding an optical vector soliton followed by classical dispersion management \cite{Tsang2006}.

Vectorial solitons in the time domain can develop an almost perfect negative correlation between quantum fluctuations around an incoherently coupled soliton pair \cite{lee2005}. The time-domain correlation patterns associated with both intrapulse and interpulse correlations have a characteristic four-domain structure. Within two of the four domains the correlations are positive, whereas for the other two domains they are negative. 

Polarization squeezed light can be generated in optical fibers driven by femtosecond pulses \cite{Andrianov2023}. The optimum squeezing ratio depends on the pulse energy and pulse duration. At short propagation distances the best squeezing was observed at energies of about 10\% larger than the fundamental soliton energy. For large distances the best squeezing ratio was achieved for pulses very close to fundamental solitons.

The interference of two bright polarization-squeezed beams on a beam splitter generates a pair of polarization-entangled light beams \cite{Korolkova2002}. An analogous scheme can be implemented in the time domain using a fiber with dispersion modulation along the fiber length. Kerr nonlinearity in optical fibers provides soliton squeezing. The dispersion modulation acts on solitons as a time-domain pulse splitter. A second-order soliton can be split into two distinct pulses propagating at different group velocities \cite{Sysoliatin2020}. The reverse process is also possible, namely, two colliding solitons can merge into a high-intensity pulse \cite{Konyukhov2020}. 

In this study, I use a linearized approach \cite{haus1990,lai1993,Rand2005} to calculate the photon-number correlations in a pulse pair generated via dispersion modulation and polarization mode dispersion. Papers \cite{haus1990,lai1993} give scalar linearized quantum theory of the soliton propagation. Vector theory was introduced in \cite{Rand2005}. 
I consider photon-number correlations in a pulse pair generated via dispersion modulation and polarization mode dispersion (PMD). Sec. \ref{sec2} reviews the linearized quantum theory of the vector soliton propagation. I consider polarization diversity homodyne detection scheme \cite{Yuan2008, Kikuchi2010, Rand2005}. This scheme allows tracking both the correlation within the selected polarization component and the cross-correlation between $x$ and $y$ polarization components of the electric field. In Sec. \ref{sec3}, I consider the pulse propagation governed by the Manakov equations \cite{manakov1973, menyuk2006, Rand2005}. 
Quantum correlations arise in the process of the splitting of second-order soliton and collision between two pulses. In Sec. \ref{sec4}, I analyze the quantum correlations in a pulse pair generated by PMD in a birefringent fiber. Sec. \ref{sec5} contains conclusions.

\section{Linearized quantum theory of pulse propagation in a birefringent fiber \label{sec2}}
 
I consider the pulse propagation in a variable-diameter birefringent fiber, 
which supports two orthogonally polarized modes. It is assumed that the modes have the same spatial profile but different propagation constants.
In the local-mode approach \cite{Snyder1983} the polarization components  of electric field can be written as follows
\begin{eqnarray}
E_k(x,y,z,t)&&=\frac{1}{2}  F(x,y) A_k(z,t) \nonumber \\
&&\exp\left( i\int_0^z\beta_{0k}(z')dz' -i \omega_0 t \right)
+\text{c.c.}, 
\end{eqnarray}

\noindent
where $k=x$ or $y$ denote the transverse coordinates,
$z$ is the propagation distance, $t$ is the time,
$F(x,y)$ is the common transverse modal profile,
$A_x$ and $A_y$ are the complex amplitudes of the polarization components of the electric field, and $\beta_{0x}$ and $\beta_{0y}$ are the propagation constants calculated for the pulse carrier frequency $\omega_0$. 
When the fiber core diameter is modulated along the fiber length, the propagation constants become $z$-dependent.

Let us introduce transformations 
\begin{eqnarray}
\tau= \frac{1}{T_0} \left( t- \int_0^z \beta_1(z') dz' \right), \quad 
\zeta=z \frac{|\overline{\beta}_2|}{T_0^2}, \label{tauzeta}     \\
U_x(\zeta,\tau)=A_x(z,t) \sqrt{F_0} \exp\left( i \int_0^z \Delta \beta(z')dz' \right), \\
U_y(\zeta,\tau)=A_y(z,t) \sqrt{F_0} \exp\left(-i \int_0^z \Delta \beta(z')dz' \right) 
\label{tauzeta2},
\end{eqnarray}

\noindent
where $T_0$ is the initial pulse width, 
$\beta_1(z)=(\beta_{1x}+\beta_{1y})/2$ is the average group delay,
$\beta_2(z)=(\beta_{2x}+\beta_{2y})/2$ is the average group velocity dispersion (GVD) parameter, $\overline{\beta}_2$ is the GVD parameter averaged over the fiber length, 
$F_0=c n \epsilon_0 {\cal A}_\mathrm{eff} T_0^2 \gamma/|\overline{\beta}_2|$ is the normalizing parameter, $c$ is the vacuum light velocity, 
$n=(c \omega_0^{-1})(\beta_{0x}+\beta_{0y})/2$ is the effective refractive index,  
$\epsilon_0$ is the vacuum permittivity,  
$\cal{A}_\mathrm{eff}$ is the effective mode area, and
$\Delta \beta(z)=\beta_{0x}-\beta_{0y}$ is related to fiber birefringence.

Neglecting losses and higher-order effects that are not significant for the chosen simulation parameters, the normalized complex amplitudes $U_x$ and $U_y$ satisfy a set of two coupled NLS equations \cite{Agrawal2013}: 

\noindent
\begin{equation}
\begin{array}{rl}
\displaystyle
\frac{\partial U_x}{\partial \zeta} + 
  b_1(\zeta) \frac{\partial U_x}{\partial \tau} = &i
  \displaystyle
  \frac{D(\zeta)}{2} \frac{\partial^2 U_x}{\partial \tau^2} + i b(\zeta) U_x 
\\*[\medskipamount]
  +i ( A |U_x|^2 &+ B |U_y|^2 ) U_x +i C U_y^2 U_x^\ast,
\\*[\medskipamount]
\displaystyle
\frac{\partial U_y}{\partial \zeta} - 
  b_1(\zeta) \frac{\partial U_y}{\partial \tau} = &i
  \displaystyle
  \frac{D(\zeta)}{2} \frac{\partial^2 U_y}{\partial \tau^2} - i b(\zeta) U_y 
\\*[\medskipamount]
  +i ( A |U_y|^2 &+ B |U_x|^2 ) U_y +i C U_x^2 U_y^\ast,
\end{array}
\label{schroedinger}
\end{equation}

\noindent
where $b=T_0^2 (\Delta \beta) (2 |\overline{\beta}_2|)^{-1}$,
$b_1=T_0 (\beta_{1x}-\beta_{1y})(2 |\overline{\beta}_2|)^{-1}$, and
$D(\zeta)=-\beta_2(z)/|\overline{\beta}_2|$
are dimensionless parameters. 
The coefficients $A, B$ and $C$ determine the specific model used in the
calculations. When $b_1=b=0$, $A=B=8/9$, and $C=0$, we have the Manakov equation
model \cite{manakov1973,menyuk2006}. This model describes the pulse propagation in a
fiber with rapidly and randomly varying birefringence. 
When $b_1 \neq 0$, $b \neq 0$, $A=1$, $B=2/3$, and $C=1/3$,  
equations (\ref{schroedinger}) describe the pulse propagation in a
fiber with linear birefringence \cite{Agrawal2013}.

In quantum theory, field amplitude functions $U_k$ and $U_k^\ast$ 
are replaced by field amplitude operators $\hat{U}_k$ 
and  $\hat{U}_k^\dagger$, $k=x,y$.
Following Haus and Lai \cite{haus1990,lai1993},
I apply the linearization approximation and expand $\hat{U}_k$ around 
the classical solution $U_k$ of the NLS equations, i.e., 
\begin{equation}
\hat{U}_k(\zeta,\tau)=U_k+\hat{u}_k(\zeta,\tau)
\label{expansion}
\end{equation} 
with 
$\langle \hat{u}_k \rangle=0$.

Inserting expansion (\ref{expansion}) into Eqs. (\ref{schroedinger}),
one finds that the operators $\hat{u}_k(\zeta,\tau)$ and
$\hat{u}_k^\dagger(\zeta,\tau)$ obey linear equations (Appendix \ref{app1}). 
With the four-component vector
\begin{equation} \label{u}
\hat{\bf u}=(\hat{u}_x^\dagger,\hat{u}_x,\hat{u}_y^\dagger,\hat{u}_y)^T
\end{equation}
linear equations (\ref{u1}) and (\ref{u2}) can be written in compact form
\begin{equation}
\frac{\partial \hat{\bf u}(\zeta,\tau)}{\partial \zeta}={\bf P} \hat{\bf u}(\zeta,\tau).
\end{equation}

Suppose that classical system is described by adjoint vector 
${\bf u}^A=(u_x^A,u_x^{A\ast},u_y^A,u_y^{A\ast})^T$. 
The inner product of two vectors is defined in the usual way as
\begin{equation} \label{uAu}
\langle {\bf u}^A | \hat{\bf u} \rangle = \frac{1}{2}
\int \limits_{-\infty}^\infty ( u_x^A \hat{u}_x^\dagger + u_x^{A\ast} \hat{u}_x +
u_y^A \hat{u}_y^\dagger+u_y^{A\ast} \hat{u}_y ) d \tau.
\end{equation}

\noindent
By requiring 
$\langle {\bf u}^A | {\bf P} \hat{\bf u} \rangle = 
\langle {\bf P}^A {\bf u}^A | \hat{\bf u} \rangle$,
one can derive evolution equation for the adjoint
vector:
\begin{equation} \label{uA}
\frac{\partial {\bf u}^A(\zeta,\tau)}{\partial \zeta}=-{\bf P}^A {\bf u}^A(\zeta,\tau).
\end{equation}
The explicit from of  Eq.~(\ref{uA}) is given in Appendix \ref{app1}. The minus sign on the right side of Eq. (\ref{uA}) is required for satisfying the conservation law 
$\partial \langle {\bf u}^A | \hat{\bf u} \rangle /\partial z =0.$
In the absence of external noise sources, the conservation law provides \cite{lai1995} 
\begin{equation} \label{var}
\langle {\bf u}^A(L,\tau) | \hat{\bf u}(L,\tau) \rangle =
\langle {\bf u}^A(0,\tau) | \hat{\bf u}(0,\tau) \rangle,
\end{equation}
where $L$ is the fiber length.
The value of ${\bf u}^A(0,\tau)$ is calculated using back propagation 
from $\zeta=L$ to $\zeta=0$
\begin{equation} \label{backprop}
{\bf u}^A(0,\tau)=\exp(-{\bf P}^A L) {\bf u}^A(L,\tau).
\end{equation}

\noindent
The initial field $(\zeta=0)$ is in coherent state. 
The perturbation operators satisfy the following commutation relations: 
\begin{eqnarray} \label{commutation}
{[}\hat{u}_k(0,\tau),\hat{u}_k^\dagger(0,\tau'){]}&=& 
\delta(\tau-\tau'),\nonumber \\ 
{[}\hat{u}_k(0,\tau),\hat{u}_k(0,\tau'){]}&=&
{[}\hat{u}_k^\dagger(0,\tau),\hat{u}_k^\dagger(0,\tau'){]}=0,\\
{[}\hat{u}_k(0,\tau),\hat{u}_n^\dagger(0,\tau'){]}&=&0, \quad k \neq n,
\nonumber 
\end{eqnarray}
where $\delta(\tau-\tau')$ is the Dirac delta function, $k,n=x,y$.

\noindent
The pulse at the fiber output ${\bf U}(L,\tau)=(U_x,U_x^\ast,U_y,U_y^\ast)$ 
can be reused as local oscillator (LO) 
in homodyne detection scheme (Appendix \ref{app2}).
The appropriate expression for the field of the LO is
\begin{equation} \label{theta}
{\bf f}_L = {\bf u}^A(L,\tau)=
{\bf U}(L,\tau) \exp(i \theta), 
\end{equation}
where $\theta$ is an adjustable phase. 

In the homodyne detection scheme the measurement operator is defined by the inner product
$\hat{M}=\langle {\bf u}^A(L,\tau) | \hat{\bf u}(L,\tau) \rangle$. In accordance with the conservation law (\ref{var}) 
\begin{equation} \label{M}
\hat{M}=\langle {\bf u}^A(0,\tau) | \hat{\bf u}(0,\tau) \rangle.
\end{equation}
The operator
\begin{equation} \label{N}
\hat{N}=\langle {\bf f_L} | \hat{\bf u}(0,\tau) \rangle
\end{equation}
is introduced for normalizing purposes.
\noindent
With Eqs. (\ref{commutation}) and (\ref{M}), squeezing ratio can be calculated as follows:
\begin{eqnarray} \label{Ratio} 
R=\frac{\text{Var}[\hat{M}]}{\text{Var}[\hat{N}]} &=&
\frac{\int _{-\infty}^{\infty} \|{\bf u}^A(0,\tau)\|^2 d\tau}
     {\int _{-\infty}^{\infty} \|{\bf u}^A(L,\tau)\|^2 d\tau},
\end{eqnarray} 
where $\text{Var}[\dots]$ stands for variance,
${\bf u}^A(0,\tau)$ is calculated using Eq. (\ref{backprop}).
The squeezed state is characterized by $R<1$.
The optimal (minimum) value
of the squeezing ratio can be chosen by varying the parameter $\theta$ in Eq.~(\ref{theta}) \cite{lai1995}.

To calculate the photon-number correlation between different parts of the pulse,
I assume that the field of the LO is filtered by a fast shutter.
For the simplicity, I consider the rectangular filter 
$Q(\tau-\tau_i)=1$ for $|\tau-\tau_i| < \Delta \tau/2$, 
and $Q(\tau-\tau_i)=0$ otherwise. 

In order to study the correlation in individual polarization components, one can
apply the backpropagation method for ${\bf U}_{x}(L,\tau)=(U_x,U_x^\ast,0,0)$ or
${\bf U}_{y}(L,\tau)=(0,0,U_y,U_y^\ast)$ function. Taking into account 
time-domain filtering, equation (\ref{theta}) can be rewritten as follows:
\begin{equation} \label{theta2}
{\bf f}_{Lk}(\tau_i) = {\bf u}^A_k(L,\tau,\tau_i)={\bf U}_k(L,\tau)
Q(\tau-\tau_i) e^{i\theta}, 
\end{equation}
where $k=x,y$.
The measurement operator corresponging to the $i$-th time slot is
\begin{eqnarray} \label{Mk}
\hat{\cal M}_k(\tau_i)=&&
    \langle {\bf u}^A_k(L,\tau,\tau_i) | \hat{\bf u}(L,\tau) \rangle 
\nonumber \\
=&& \langle {\bf u}^A_k(0,\tau,\tau_i) | \hat{\bf u}(0,\tau) \rangle.
\end{eqnarray}
The normalizing operator is
\begin{equation} \label{Nk}
\hat{\cal N}_{k}(\tau_i)=\langle {\bf f}_{Lk}(\tau_i) | \hat{\bf u}(0,\tau) \rangle.
\end{equation}

\noindent
The photon-number correlations between two time slots is given by the normally ordered covariance
\begin{eqnarray} \label{Correlxxyy}
&&C_{kn}(\tau_i,\tau_j)= \nonumber \\
&&\frac{
\langle: \hat{\cal M}_k(\tau_i) \hat{\cal M}_n(\tau_j)  :\rangle 
}
{\left( \langle \hat{\cal N}_k^\dagger(\tau_i) \hat{\cal N}_k(\tau_i) \rangle 
\langle \hat{\cal N}_n^\dagger(\tau_j) \hat{\cal N}_n(\tau_j) \rangle \right)^{1/2}
}=\\
&&
\frac{ \langle {\bf u}_k^A(0,\tau,\tau_i) | {\bf u}_n^{A \ast}(0,\tau,\tau_i) \rangle 
      -\delta_{kn}\delta(\tau_i-\tau_j)}
{\left( \int _{-\infty}^{\infty} \|{\bf u}_k^A(L,\tau,\tau_i)\|^2 d\tau
        \int _{-\infty}^{\infty} \|{\bf u}_n^A(L,\tau,\tau_j)\|^2 d\tau \right)^{1/2}},
\nonumber
\end{eqnarray}
where $k,n=x,y$, and ${\bf u}_k^A(0,\tau,\tau_i)$ is calculated from ${\bf u}_k^A(L,\tau,\tau_i)$ using back propagation (\ref{backprop}).

Now the complete correlation function $C$ can be introduced. With $\hat{\cal M}(\tau_i)=\hat{\cal M}_x(\tau_i)+\hat{\cal M}_y(\tau_i)$ and $\hat{\cal N}(\tau_i)=\hat{\cal N}_x(\tau_i)+\hat{\cal N}_y(\tau_i)$ the complete correlation function can be written as follows
\begin{eqnarray} \label{Correl}
&&C(\tau_i,\tau_j)= \nonumber \\
&&\frac{
\langle: \hat{\cal M}(\tau_i) \hat{\cal M}(\tau_j)  :\rangle
}
{\left( \langle \hat{\cal N}^\dagger(\tau_i) \hat{\cal N}(\tau_i) \rangle 
\langle \hat{\cal N}^\dagger(\tau_j) \hat{\cal N}(\tau_j) \rangle \right)^{1/2}
}= \\
&&\frac{ 
\langle {\bf u}^A(0,\tau,\tau_i) | {\bf u}^{A \ast}(0,\tau,\tau_i) \rangle 
      -\delta(\tau_i-\tau_j)}
{\left( \int _{-\infty}^{\infty} \|{\bf u}^A(L,\tau,\tau_i)\|^2 d\tau
        \int _{-\infty}^{\infty} \|{\bf u}^A(L,\tau,\tau_j)\|^2 d\tau \right)^{1/2}}.
\nonumber
\end{eqnarray}
The functions  $C_{xx}$ and $C_{yy}$ (\ref{Correlxxyy}) give quantum noise correlations in a selected polarization component, $U_x$ or $U_y$. The function $C_{xy}=C_{yx}$ gives the cross-correlations of the noise in the polarization components $U_x$ and $U_y$. The complete correlation function $C(\tau_i,\tau_j)$ shows both the correlations in a selected polarization component and the cross-correlations of the noise in two polarization components.

Instead of time-domain filter in (\ref{theta2}) one can use frequency domain narrow-bandpass filter, e.g., $W(\Omega-\Omega_i)=1$ for $|\Omega-\Omega_i| < \Delta \Omega/2$, and $W(\Omega-\Omega_i)=0$ otherwise. Here $\Omega=(\omega-\omega_0) T_0$ is the normalized frequency difference. With the frequency-domain filtering the equation for adjoint field (\ref{theta}) can be written as follows
\begin{equation} \label{theta3}
{\bf u}^A_k(L,\tau,\Omega_i) =
{\cal F}^{-1} \Bigl[ W(\Omega-\Omega_i)
{\cal F} \bigl[ {\bf U}_k(L,\tau) \bigr] \Bigr] e^{i\theta}
\end{equation}
where ${\cal F}$ and ${\cal F}^{-1}$ denotes direct and inverse Fourier transform. Following the procedure described above, one can calculate photon-number correlations in the frequency domain:
${S}_{xx}(\Omega_i,\Omega_j)$, ${S}_{yy}(\Omega_i,\Omega_j)$,
${S}_{xy}(\Omega_i,\Omega_j)$, and ${S}(\Omega_i,\Omega_j)$.

\section{Manakov model \label{sec3}}

With $b_1=b=0$, $A=B=8/9$, and $C=0$ Eqs.~(\ref{schroedinger}) are known as the Manakov equations \cite{manakov1973}. The Manakov equations are solved numerically with an input in the form 
\begin{equation} \label{Uini}
U_x(0,\tau)=U_y(0,\tau)=\frac{1}{\sqrt{2}}u_0\,\text{sech}(\tau). 
\end{equation}
I choose the input pulse polarization to be linear so that the $x$ and $y$ components of the pulse are equally intense. Parameter $u_0$ is related to the peak power of input pulses. 
In the scalar approach, a fundamental soliton is formed for $u_0$ in the range of 0.5 to 1.5 \cite{Agrawal2013}.

\begin{figure}
\begin{tabular}{ll}
(a) & (b)\\
\includegraphics[height=1.1in]{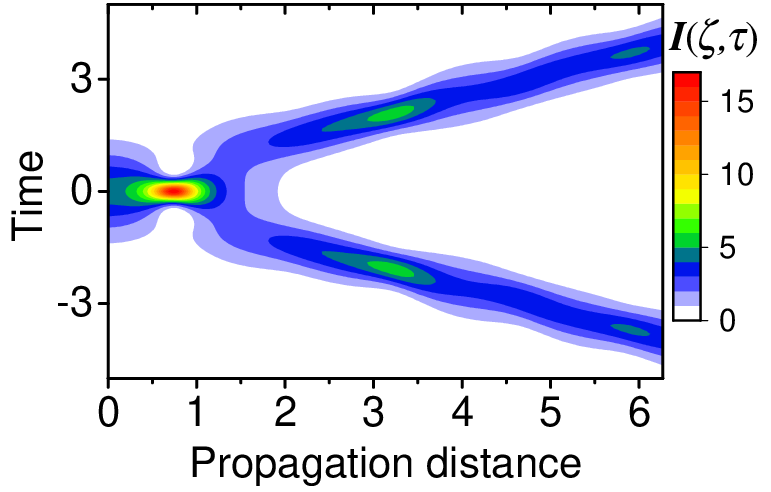} &
\includegraphics[height=1.1in]{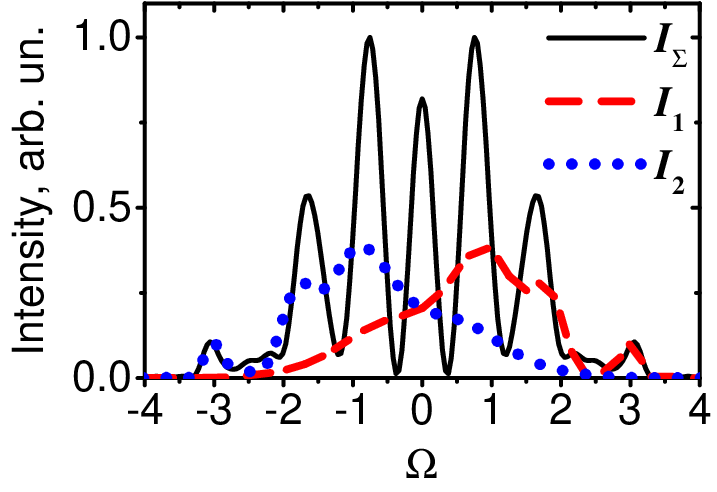}
\end{tabular}
\begin{tabular}{lll}
(c) & (d) & (e)\\
\includegraphics[height=1.2in]{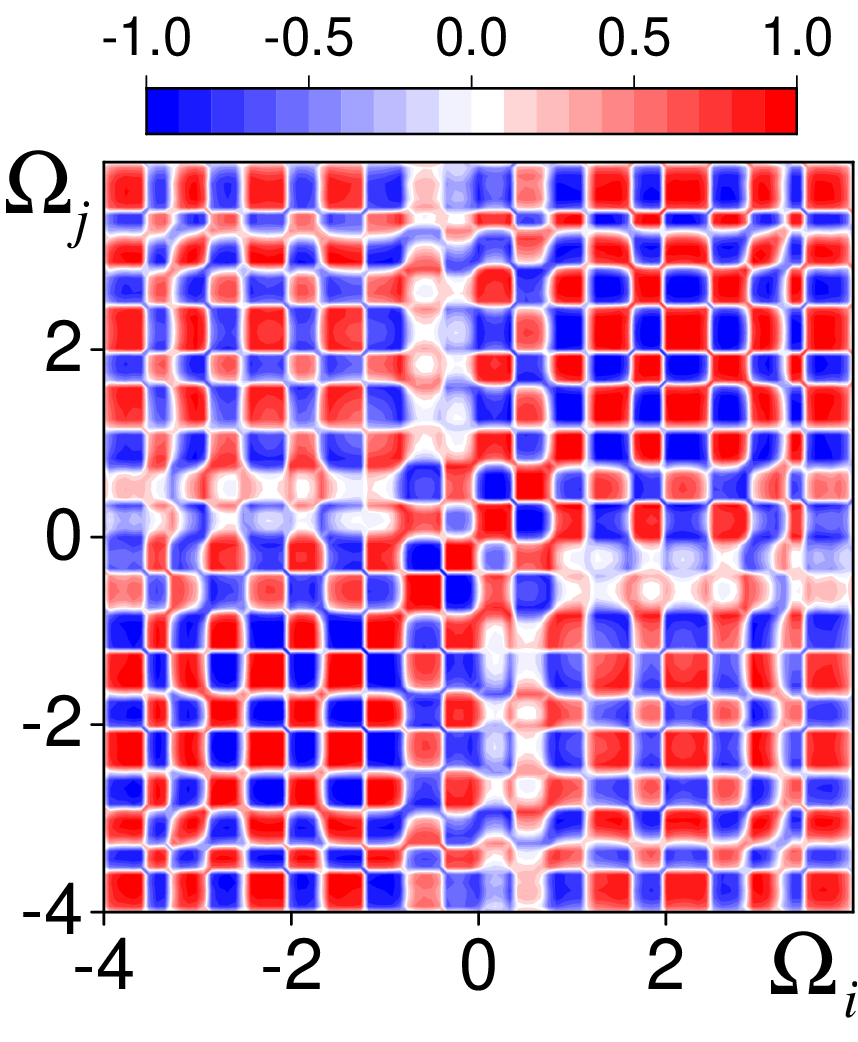}  & 
\includegraphics[height=1.2in]{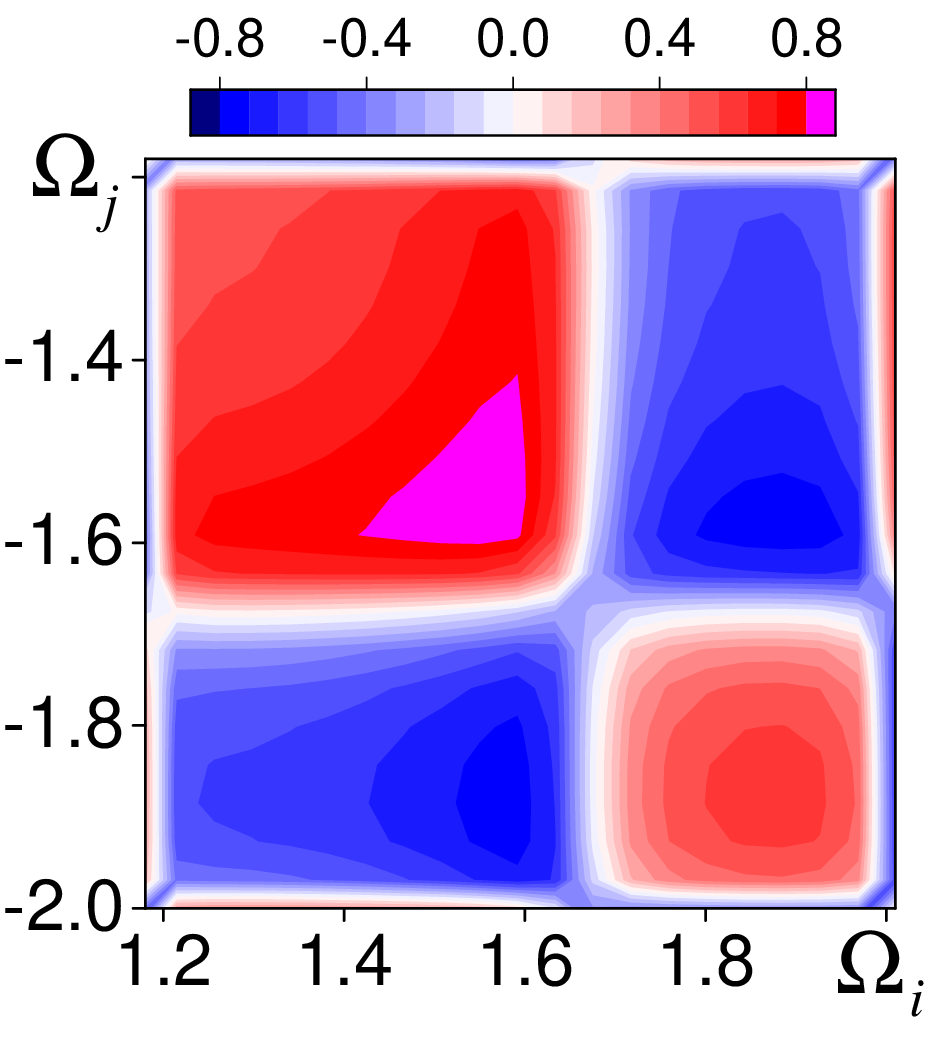}  &
\includegraphics[height=1.2in]{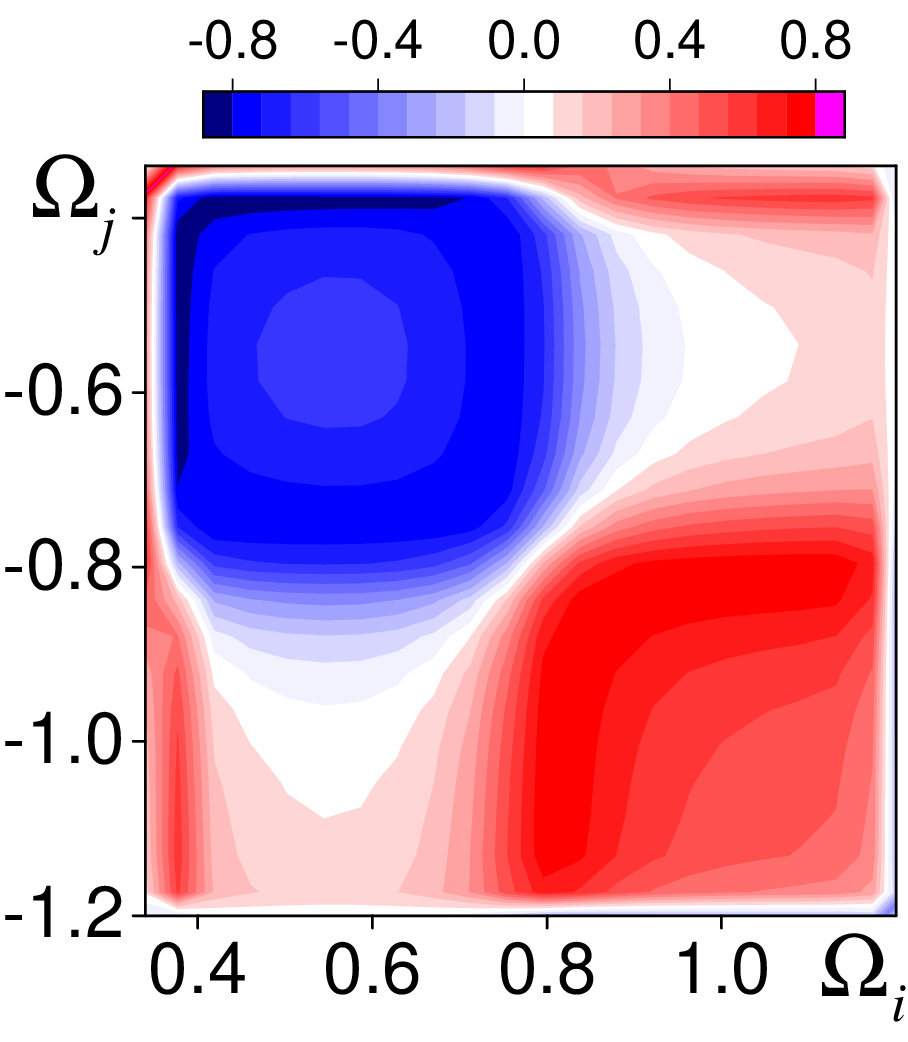}  
\end{tabular}
\caption{
Splitting of the second-order soliton. 
(a) Temporal evolution of the pulse field intensity $I(\zeta,\tau)=|U_x|^2+|U_y|^2$. 
(b) Output spectrum $I_\Sigma(L,\Omega)$ (solid curve) and spectra of the first $I_1(L,\Omega)$ (dashed curve) and second $I_2(L,\Omega)$ (dotted curve) pulses. 
(c), (d), and (e) The correlation function $S(\Omega_i,\Omega_j)$ calculated at $\zeta=L$ for three different frequency regions. 
Simulation parameters are $u_0=2$, $\zeta_m=1.3$, and $L=2\pi$.
}
\label{fig1}
\end{figure}

Figure \ref{fig1}(a) shows the splitting of a second-order soliton in a fiber with dispersion modulation 
\begin{equation} \label{sine}
D(\zeta)=1-0.2\sin(2\pi \zeta/\zeta_m),
\end{equation}
where $\zeta_m$ is the modulation period. The initial pulse is polarized linearly. After splitting, the pulses maintain initial polarization so that the $x$ and $y$ components of the pulses are equally intense. The pulses have opposite frequency shift and propagate with different group velocities. At the output $(\zeta=2\pi)$ the pulses overlap weakly, and their spectra can be calculated separately. 
In Fig.~\ref{fig1}(b) the spectra of the first pulse and second one are $I_1(L,\Omega)=|{\cal F}[(1-H(\tau))U_x]|^2+|{\cal F}[(1-H(\tau))U_y]|^2$ and $I_2(L,\Omega)=|{\cal F}[H(\tau)U_x]|^2+|{\cal F}[H(\tau)U_y]|^2$, respectively. 
Here $H(\tau)$ is the Heaviside step function, ${\cal F}[\dots]$ stands for Fourier transform. The spectra of the output pulses overlap. The interference between two frequency-shifted pulses causes the modulation of the output spectrum $I_\Sigma(L,\Omega)=|{\cal F}[U_x]|^2+|{\cal F}[U_y]|^2$ [Fig.~\ref{fig1}(b)]. The function $I_\Sigma(L,\Omega)$  contains seven distinct local maxima. Correlations arise inside and between local maxima.  
As a result, the spectral correlation function $S(\Omega_i,\Omega_j)$ acquires a checkerboard structure with alternating positive and negative regions [Fig.~\ref{fig1}(c)].
All correlation patterns are calculated at the angle $\theta$ that corresponds to the best squeezing ratio $R$ (\ref{Ratio}).

Figure \ref{fig1}(d) shows the correlation associated with two local maxima of the output spectrum. The first local maximum is at a frequency $\Omega=1.63$, and the second one is at $\Omega=-1.63$ [Fig.~\ref{fig1}(b)]. The correlation function $S(\Omega_i,\Omega_j)$ contains four quadrants [Fig.~\ref{fig1}(d)]; two of them show negative photon-number correlations and the other two off-diagonal quadrants show positive photon-number correlations.
The correlation associated with the intensity peaks at $\Omega=0.75$ and $\Omega=-0.75$ has quite another distribution [Fig.~\ref{fig1}(e)]. High values of the correlation function were obtained only in two off-diagonal quadrants. One of them shows a negative correlation with a minimum negative value of $S=-0.86$, and the other shows a positive correlation with a maximum value of $S=0.79$.

At frequencies corresponding to the local maximum of the spectrum [Fig.~\ref{fig1}(b)], the correlation function approaches zero. The absolute values of the correlation function $S(\Omega_i,\Omega_j)$ reach their maximum at the periphery of the local maximum of the spectrum. 

Fig.~\ref{fig1}(c) shows the spectral distribution of the correlation function $S(\Omega_i,\Omega_j)$. The correlation functions $S_{xx}$, $S_{yy}$ and $S_{xy}$ have simalar distributions because the output field is linearly polarized and the $x$ and $y$ components of the output pulses are equally intense.


The soliton splits into two pulses after the first modulation period [Fig.~\ref{fig1}(a)].
After splitting, the propagation of the pulses in the fiber with dispersion modulation is accompanied by the generation of low-intensity dispersive waves \cite{Hasegawa1991}.
The generation of dispersive waves can be reduced using truncated sine modulation
\begin{equation} \label{sinetr}
D(\zeta)= 
\left\{
\begin{array}{ll}
1-0.2 \sin(2\pi\zeta/\zeta_m), & 0 \leq \zeta \leq \zeta_m,\\
1,                             & \zeta > \zeta_m.
\end{array}
\right.
\end{equation}
With the modulation given by Eq.(\ref{sinetr}), the soliton splits after one modulation period. Then, both pulses propagate separately. In a fiber segment $\zeta>\zeta_m$ the modulation is absent and dispersive waves are not enhanced. 

\begin{figure}
\begin{tabular}{ll}
(a) & (b)\\
\includegraphics[height=1.1in]{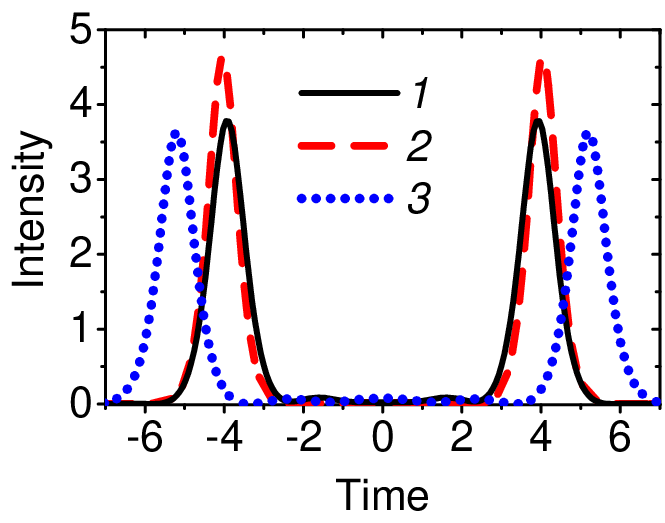} &
\includegraphics[height=1.1in]{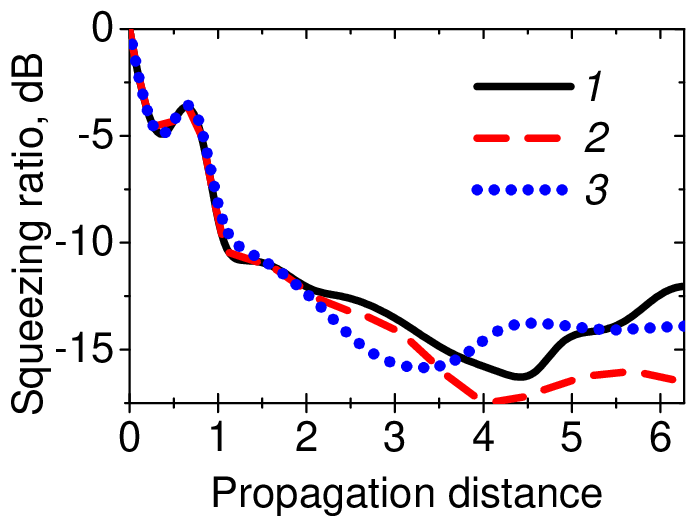}
\end{tabular}
\begin{tabular}{lll}
(c) & (d) & (e)\\
\includegraphics[height=1.2in]{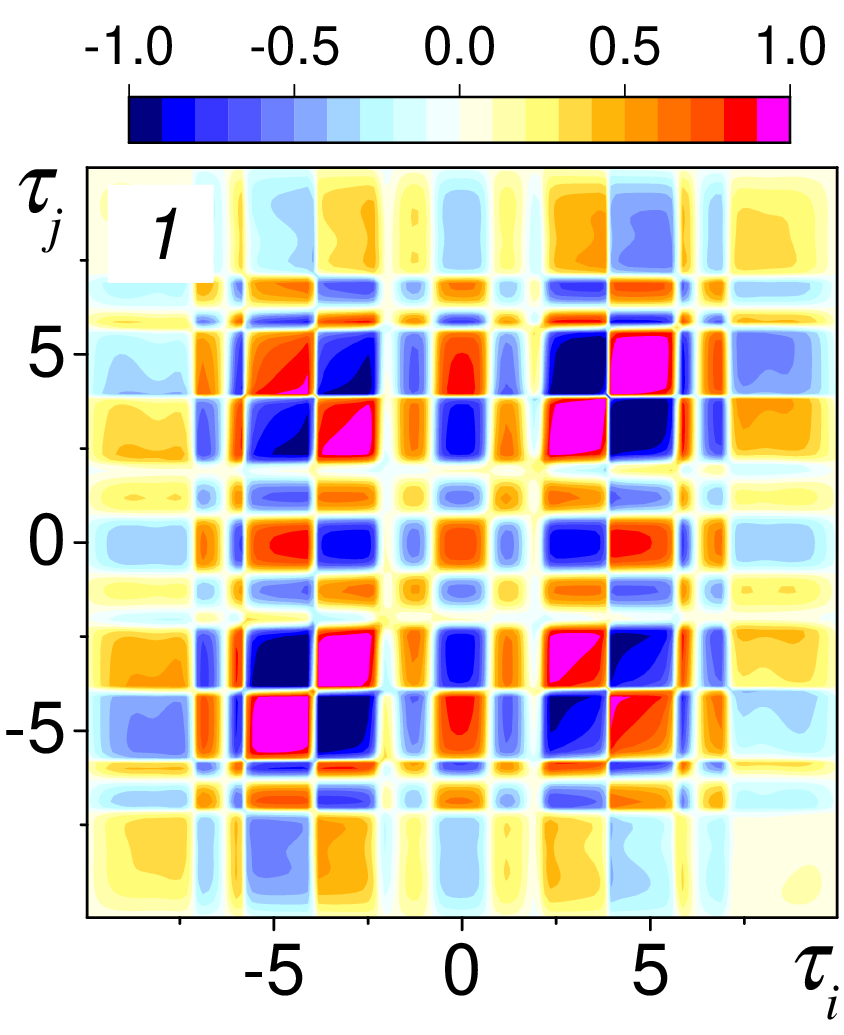}  & 
\includegraphics[height=1.2in]{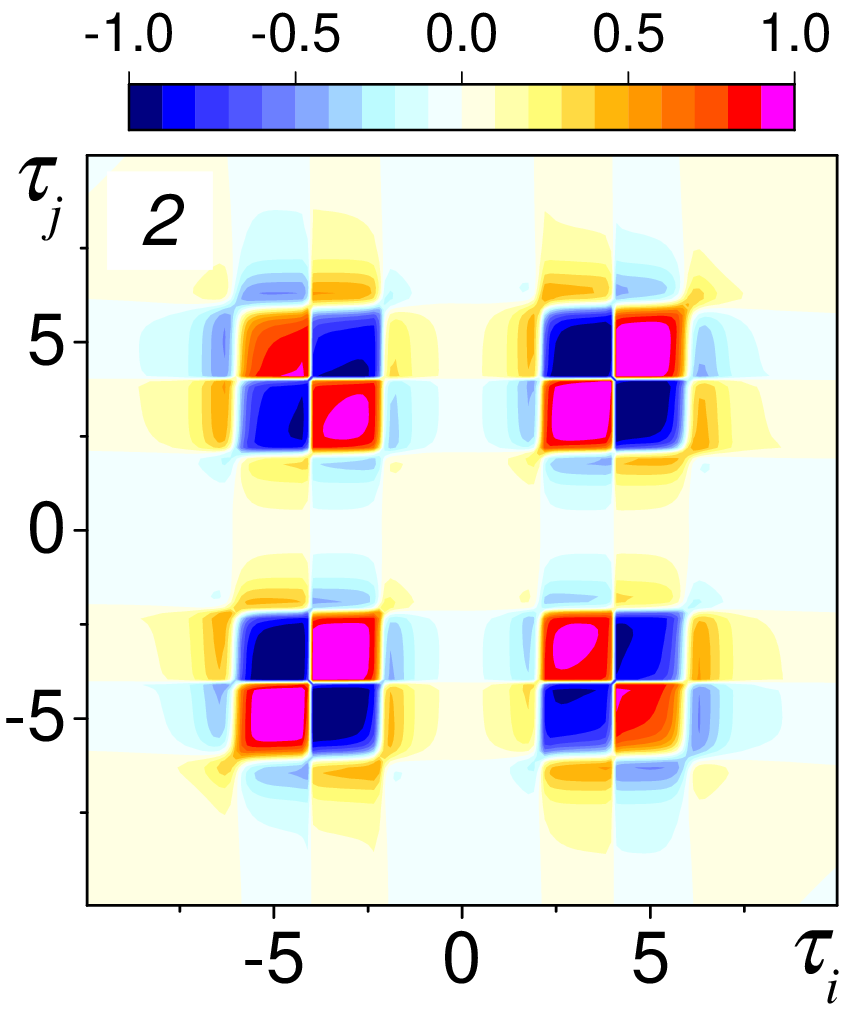}  &
\includegraphics[height=1.2in]{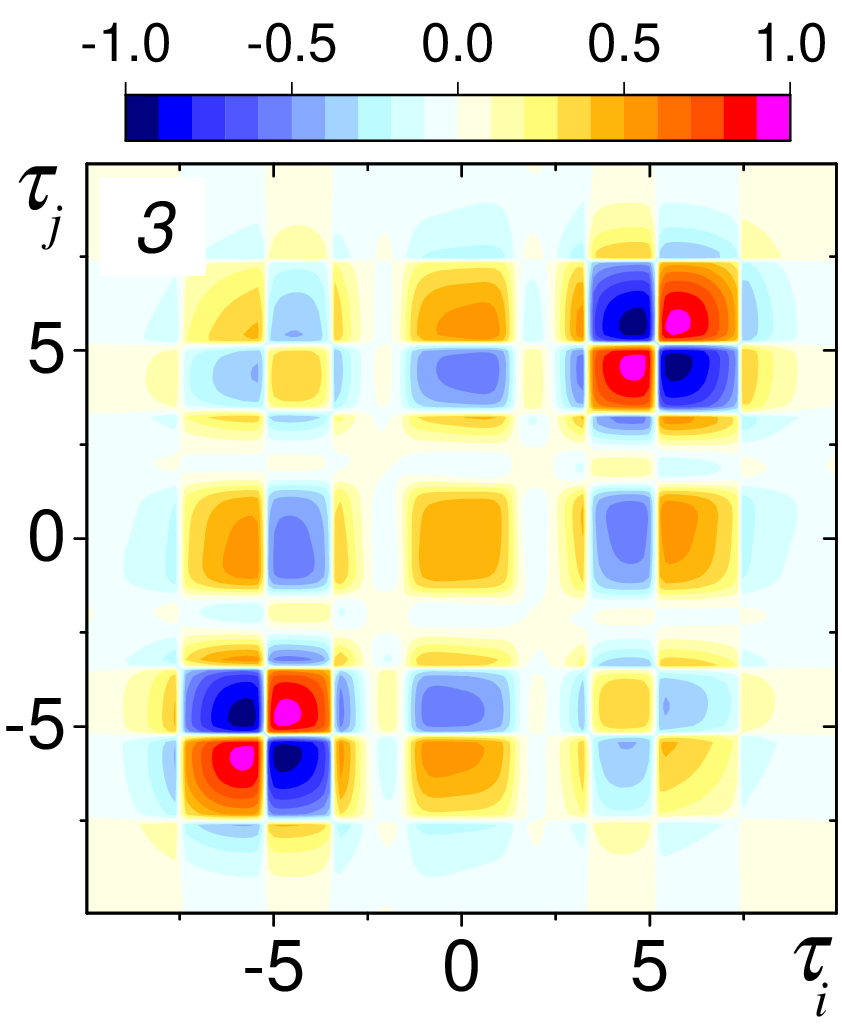}  
\end{tabular}
\caption{
Squeezing ratio and time-domain correlations for regime shown in Fig.~\ref{fig1}. 
(a) Output field intensity $I(L,\tau)=|U_x|^2+|U_y|^2$, $L=2\pi$. 
(b) The best squeezing ratio $\text{min}_\theta R$. 
(c), (d), and (e) show time-domain complete correlation function $C(\tau_i,\tau_j)$ (\ref{Correl}) calculated at a distance $\zeta=L=2\pi$. Labels \textsl{1}, \textsl{2}, and \textsl{3} correspond to three different conditions for the modulation of the fiber dispersion: \textsl{1} corresponds to the sine-wave modulation (\ref{sine}) at $\zeta_m=1.3$,  \textsl{2}  to truncated sine modulation (\ref{sinetr}) at $\zeta_m=1.3$, and \textsl{3} to the sine-wave modulation (\ref{sine}) at $\zeta_m=\pi/2$.
Other parameters are the same as in Fig.~\ref{fig1}. 
}
\label{fig2}
\end{figure}

Figure \ref{fig2}(a) shows the output pulse profiles. Curves \textsl{1} and \textsl{2} are calculated for the same modulation period but different modulation profiles, namely sine-wave modulation (\ref{sine}) and truncated sine modulation (\ref{sinetr}).  Sine-wave modulation reduces pulse peak intensities as the part of the pulse energy is radiated to dispersive waves. Figure \ref{fig2}(b) shows the squeezing ratio obtained with both sine-wave modulation (\ref{sine}) and truncated sine modulation (\ref{sinetr}). Curve \textsl{3} in Fig.~\ref{fig2}(a) is calculated at a resonance condition when the soliton period coincides with the modulation period $\zeta_m=\pi/2$ \cite{Hasegawa1991}. The resonance condition provides maximum separation between the peaks of the output pulses. 

At the output $(\zeta=2\pi)$ the best squeezing is achieved using truncated sine modulation, see curve \textsl{2} in Fig.~\ref{fig2}(b). With sine-wave modulation the squeezing ratio degrades, as shown by curves \textsl{1} and \textsl{3} in Fig.~\ref{fig2}(b). 

Changing the modulation period or modulation type significantly affects the time-domain correlation pattern. Figures \ref{fig2}(c)-(e) show the complete correlation function $C(\tau_i,\tau_j)$ both for sine-wave modulation and truncated sine modulation. Other correlation functions $C_{xx}$, $C_{yy}$, and $C_{xy}$ have similar distributions. The correlation pattern is similar to that obtained for scalar solitons \cite{lee2005}. The coordinate plane of the function $C(\tau_i,\tau_j)$ can be divided into four quadrants. The two diagonal quadrants represent intrapulse photon-number correlations. The other two off-diagonal quadrants represent interpulse correlations. 

With sine-wave modulation, photon-number correlations arise not only between pulses but also between low-intensity dispersive waves [Fig.~\ref{fig2}(c)]. With truncated sinusoidal modulation, the intensity of dispersive waves is significantly reduced. As a result, the correlation pattern is free of fringes associated with dispersive waves [Fig.~\ref{fig2}(d)]. Changing the modulation period to the resonance value $\zeta_m=\pi/2$ increases the time separation of the pulses [Fig.~\ref{fig2}(a)]. However, with $\zeta_m=\pi/2$ the interpulse correlations disappear [Fig.~\ref{fig2}(e)]. This example shows that different conditions are required for the best pulse separation and the formation of highly correlated pulse pairs.

Now I turn to the case of colliding solitons. The initial field is given by
\begin{equation} \label{Uini2}
\begin{array}{rl}
U_x(0,\tau)&=
\displaystyle
\frac{1}{\sqrt{2}}u_0\,\text{sech}(\tau+T)\exp(i\Delta\omega\,\tau), 
\\*[\bigskipamount] 
U_y(0,\tau)&=
\displaystyle
\frac{1}{\sqrt{2}}u_0\,\text{sech}(\tau-T)\exp(-i\Delta\omega\,\tau),
\end{array}
\end{equation}
where $T$ defines the temporal distance between initial pulses and $\Delta\omega$ is initial frequency shift.

\begin{figure}
\begin{tabular}{l}
(a) \\
\includegraphics[width=1.2in]{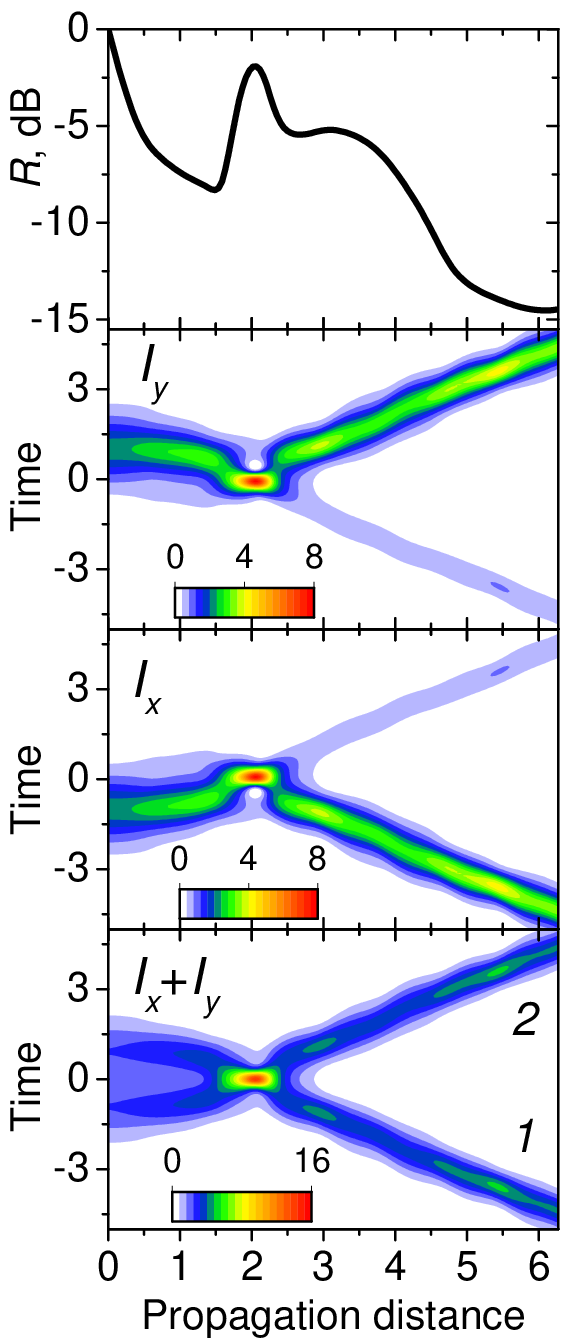}
\end{tabular}
\begin{tabular}{ll}
(b) & (c)\\
\includegraphics[width=1.0in]{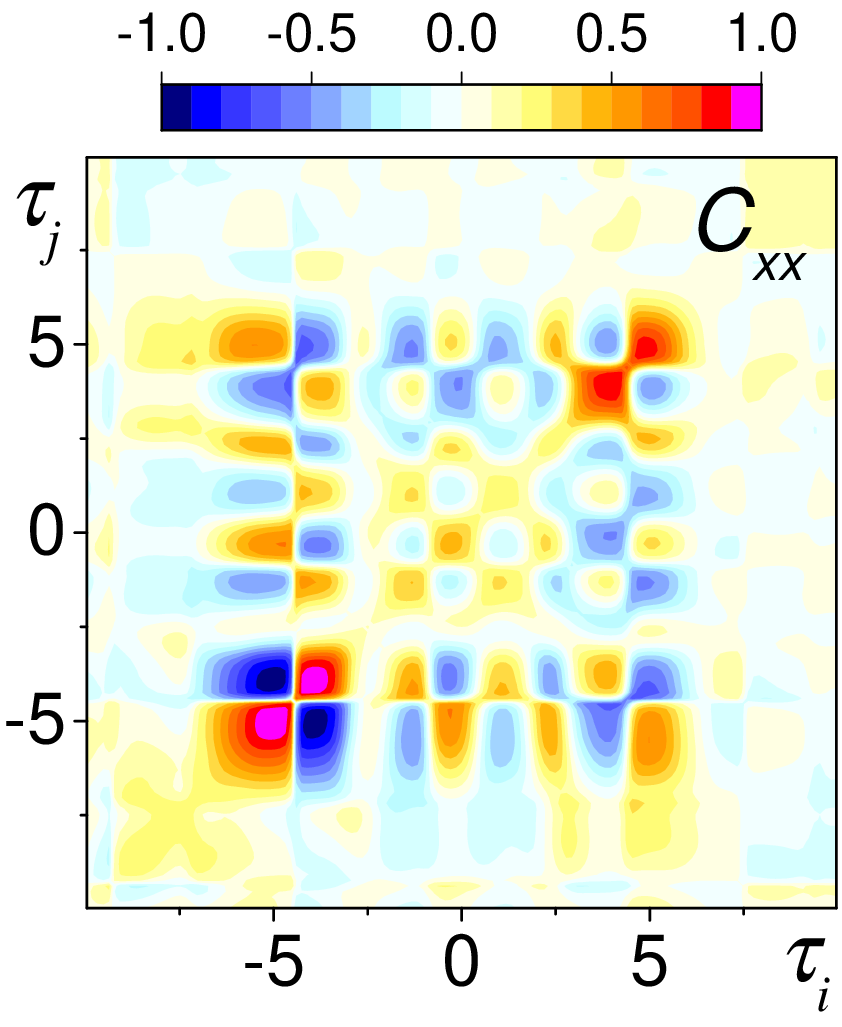} & \includegraphics[width=1.0in]{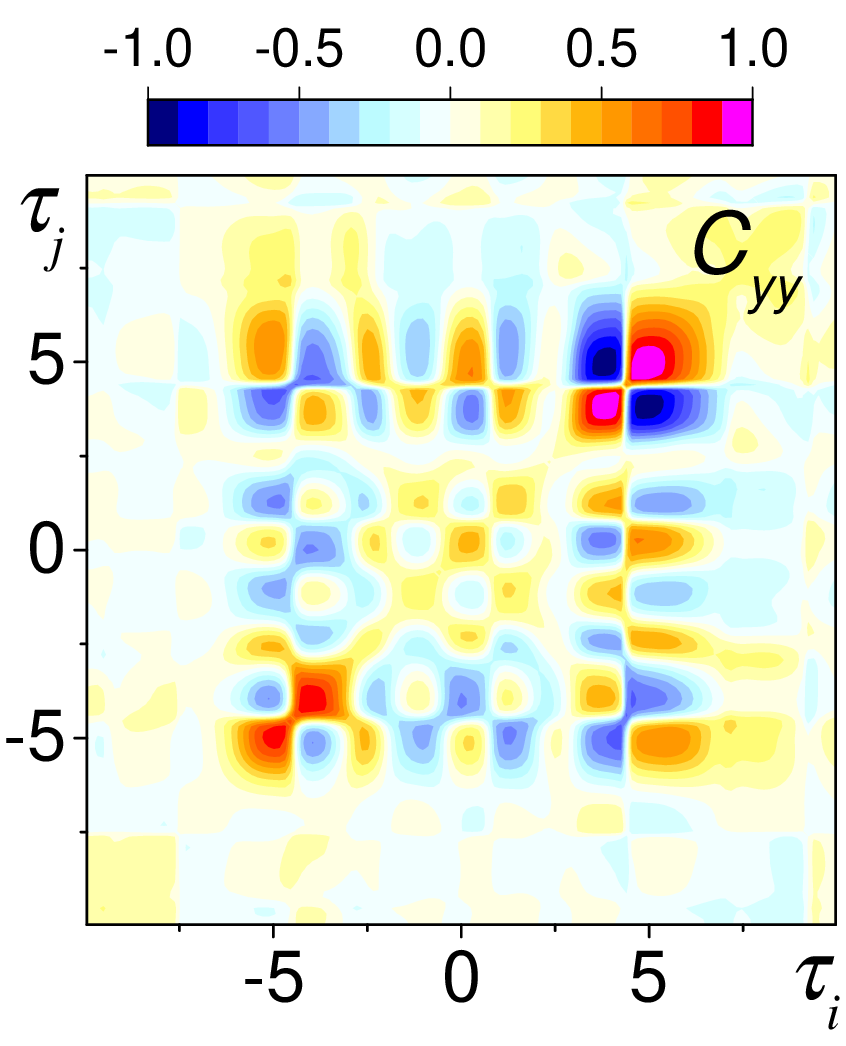}\\
(d) & (e)\\
\includegraphics[width=1.0in]{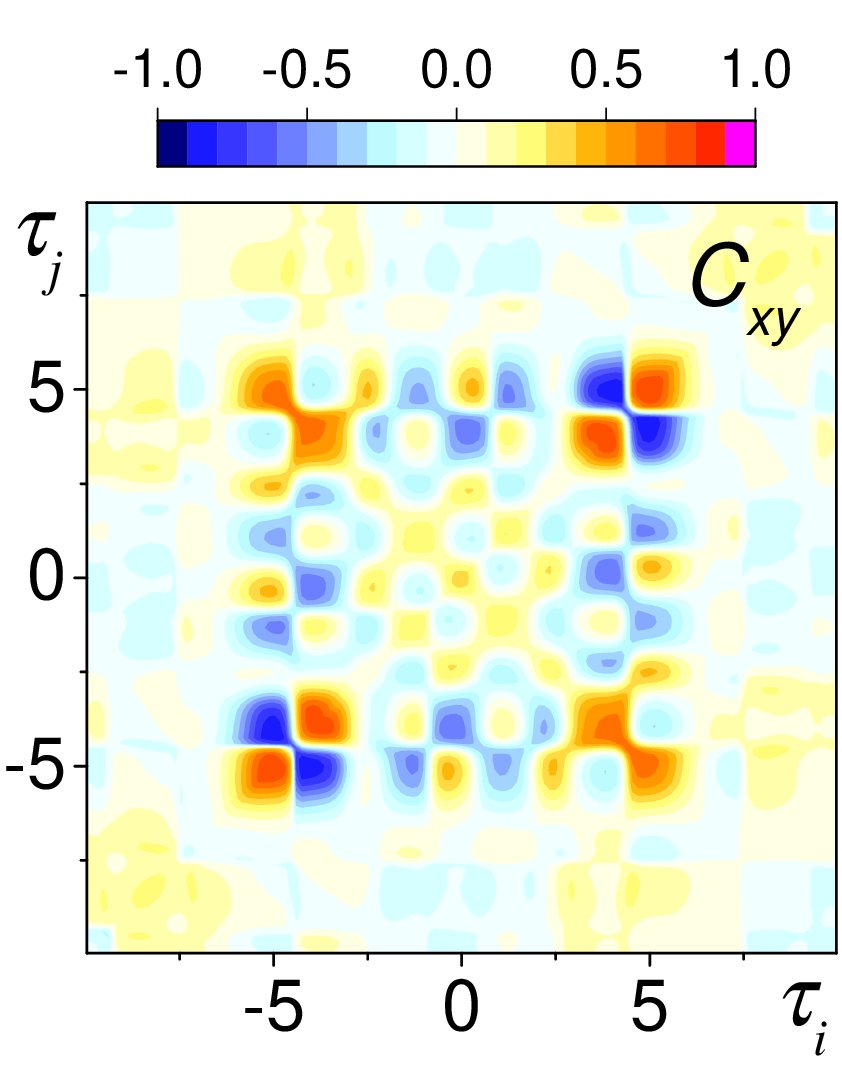} & \includegraphics[width=1.0in]{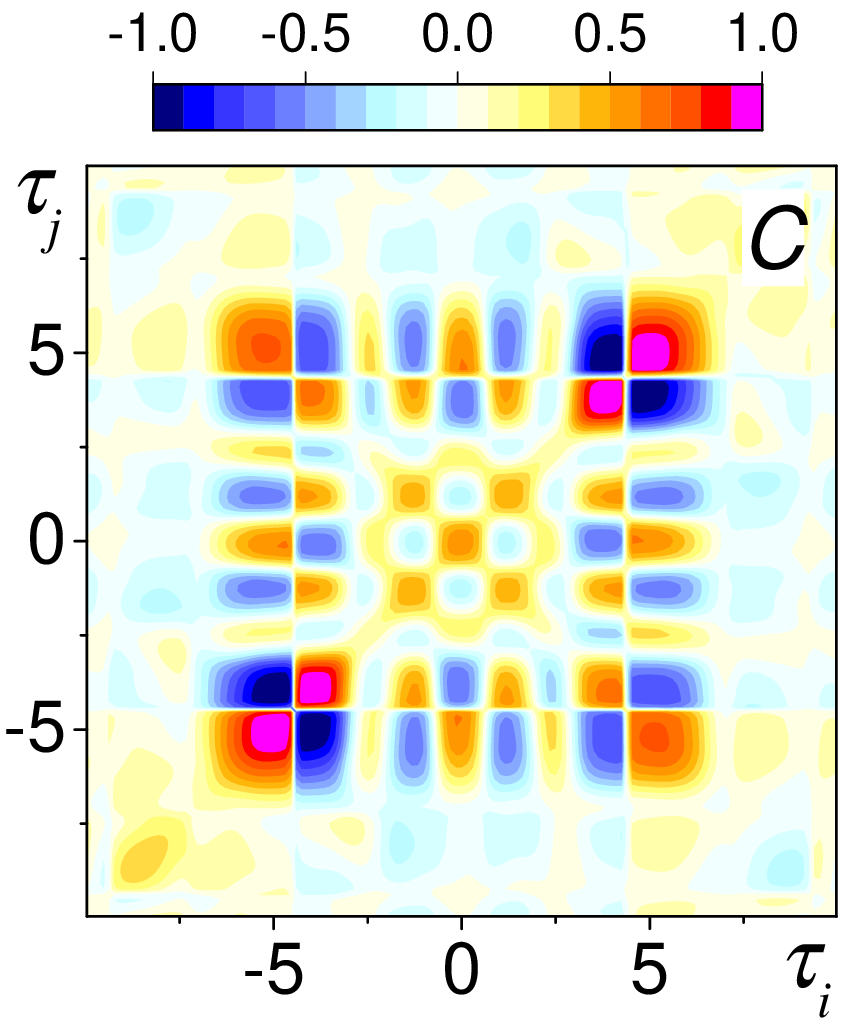}
\end{tabular}
\caption{ \label{fig3}
Inelastic collision of two co-propagating solitons. 
(a) From top to bottom: the best squeezing ratio $\text{min}_\theta R$, intensity of the $y$-component $I_y(\zeta,\tau)=|U_y|^2$, intensity of the $x$-component $I_x(\zeta,\tau)=|U_x|^2$, and total intensity $I(\zeta,\tau)=I_x+I_y$. 
(b), (c), (d), and (e) show time-domain correlation functions calculated at a distance $\zeta=L=2\pi$. Other parameters are $u_0=2$, $T=1$, $\Delta \omega=0$, and $\zeta_m=0.83$. }
\end{figure}

With $\Delta \omega=0$ in Eq. (\ref{Uini2}) the initial pulses have the same central frequency and group velocity. In the absence of modulation, i.e., at $\zeta_m=\infty$ in Eq.~(\ref{sine}), the pulses attract and repel periodically. The modulation violates periodicity in the propagation of the pulses. With dispersion modulation the pulses collide and then reflect off each other [Fig.~\ref{fig3}(a)]. After reflecting, the propagation of the pulses is accompanied by the enhancement of the squeezing ratio, see the top part of Fig.~\ref{fig3}(a). 

In the bottom part of Fig.~\ref{fig3}(a) the first and second output pulses are denoted by \textsl{1} and \textsl{2}. The output field has elliptical polarization, which varies across the pulses. The contribution of the $I_x=|U_x|^2$ and $I_y=|U_y|^2$ components to the intensity $I=|U_x|^2+|U_y|^2$ of the first and second pulses is different. 
As a result, the correlation function $C_{xx}$ has peak values $\pm 1$ in the vicinity of the point $(\tau_i,\tau_j)=(-4.5,-4.5)$ [Fig.~\ref{fig3}(b)], which corresponds to the peak intensity of the first pulse. The correlation function  $C_{yy}$ has peak values about the point $(\tau_i,\tau_j)=(4.5,4.5)$ [Fig.~\ref{fig3}(c)] associated with the second pulse. The points $(\tau_i,\tau_j)=(\pm 4.5,\pm 4.5)$ correspond to the intrapulse correlations. Strong intrapulse correlations can be found both in the $C_{xx}$, $C_{yy}$ and $C_{xy}$ functions [Fig.~\ref{fig3}(d)]. 

The interpulse correlations can be found in the vicinity of points $(\tau_i,\tau_j)=(4.5,-4.5)$ and $(\tau_i,\tau_j)=(-4.5,4.5)$. The relationship between the intrapulse and interpulse correlations can be seen from Fig.~\ref{fig3}(e), which shows the complete correlation function $C(\tau_i,\tau_j)$. The interpulse correlations have peak values $\pm 0.5$, while the intrapulse correlations approach $\pm 1$.    

\begin{figure}
\includegraphics[height=1.1in]{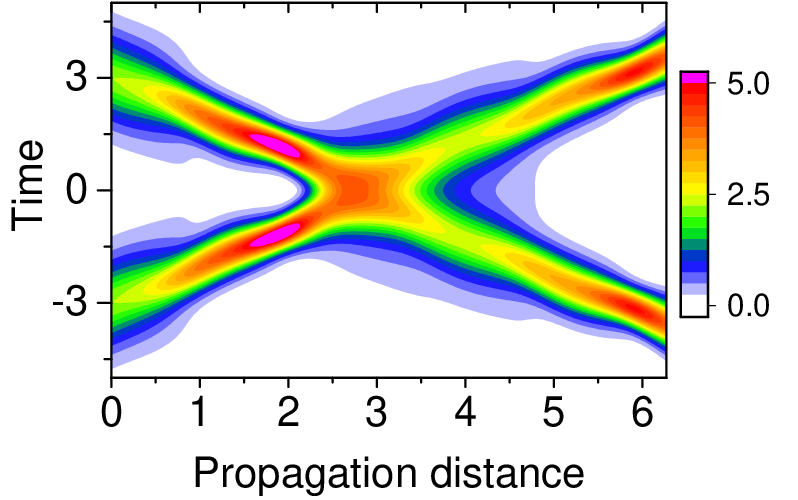}
\caption{ \label{fig4}
Collision of two frequency-shifted pulses. 
False color plot shows the field intensity $I(\zeta,\tau)=|U_x|^2+|U_y|^2$. 
Initial field is given by Eq.~(\ref{Uini2}). The initial pulse parameters are $u_0=2$, $T=3$, and $\Delta \omega=1$. The modulation period in Eq. (\ref{sine}) is $\zeta_m=1.3$. }
\end{figure}

\begin{figure}
\begin{tabular}{llll}
(a) & (b) & (c) & (d)\\
\includegraphics[width=0.8in]{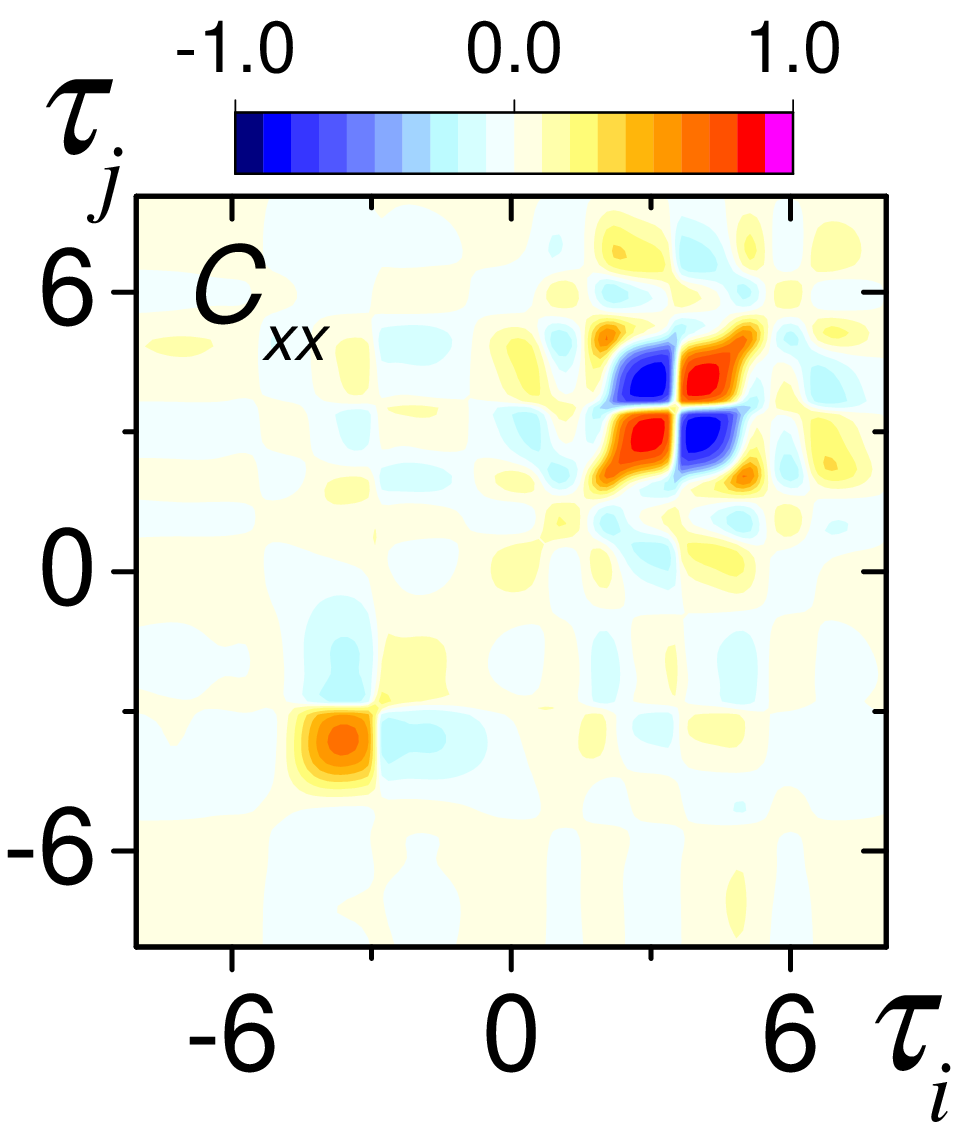} & 
\includegraphics[width=0.8in]{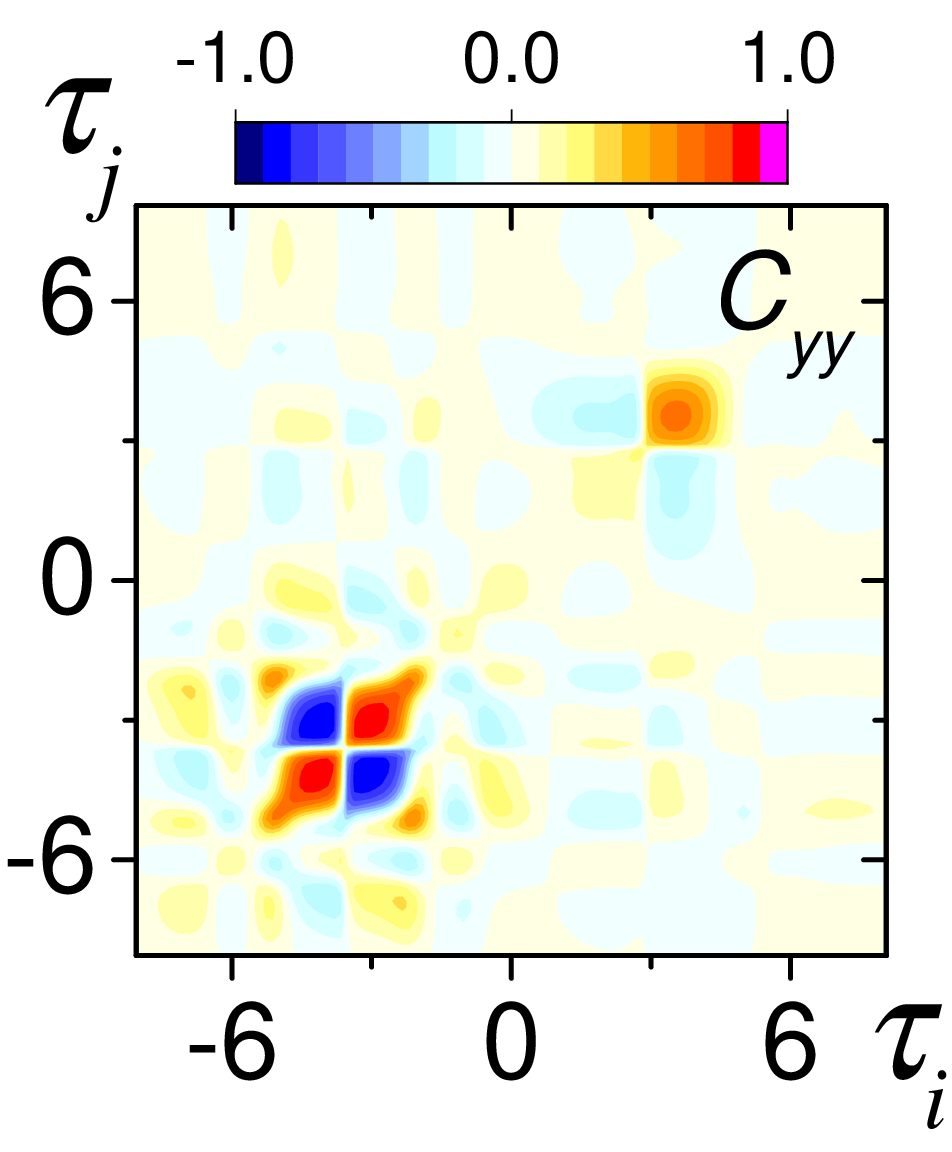} &
\includegraphics[width=0.8in]{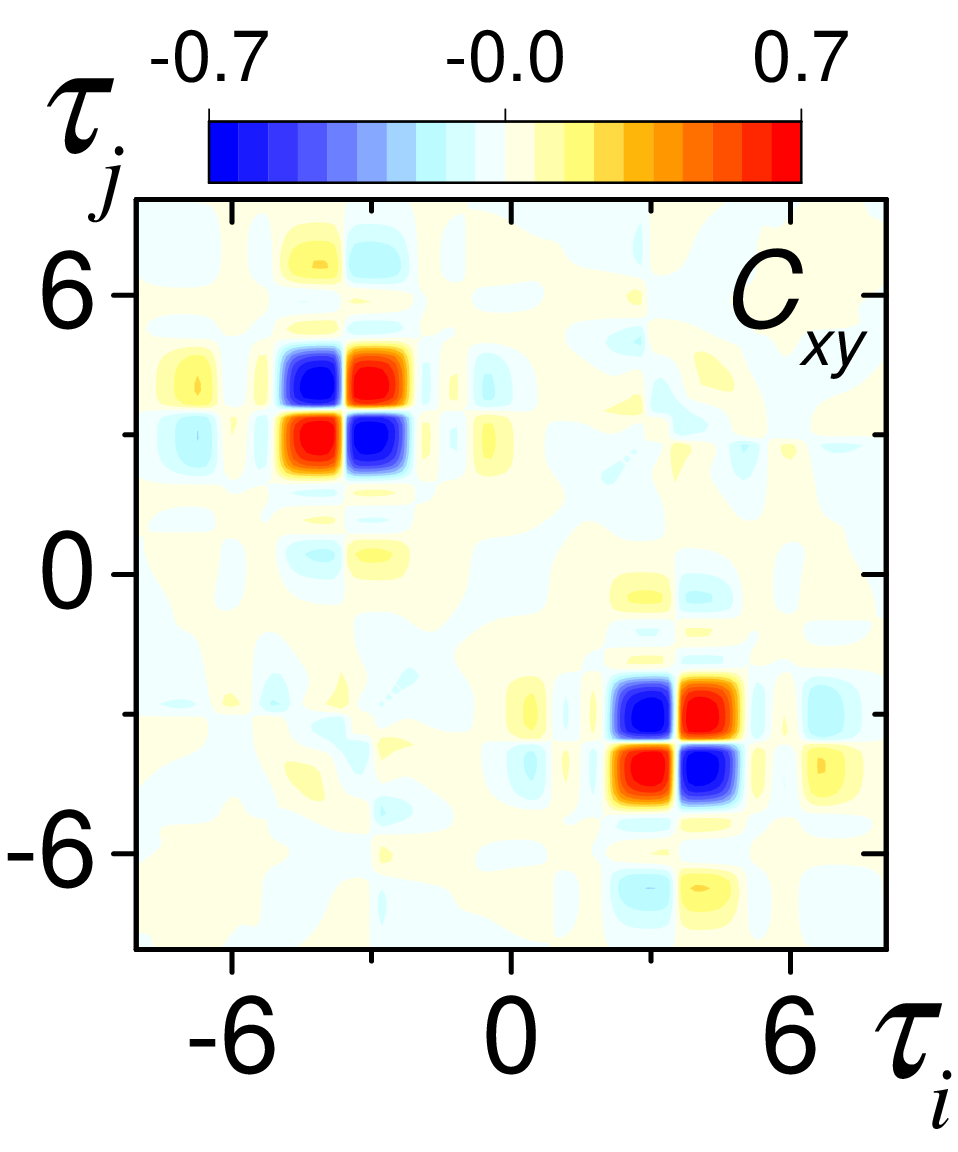} & 
\includegraphics[width=0.8in]{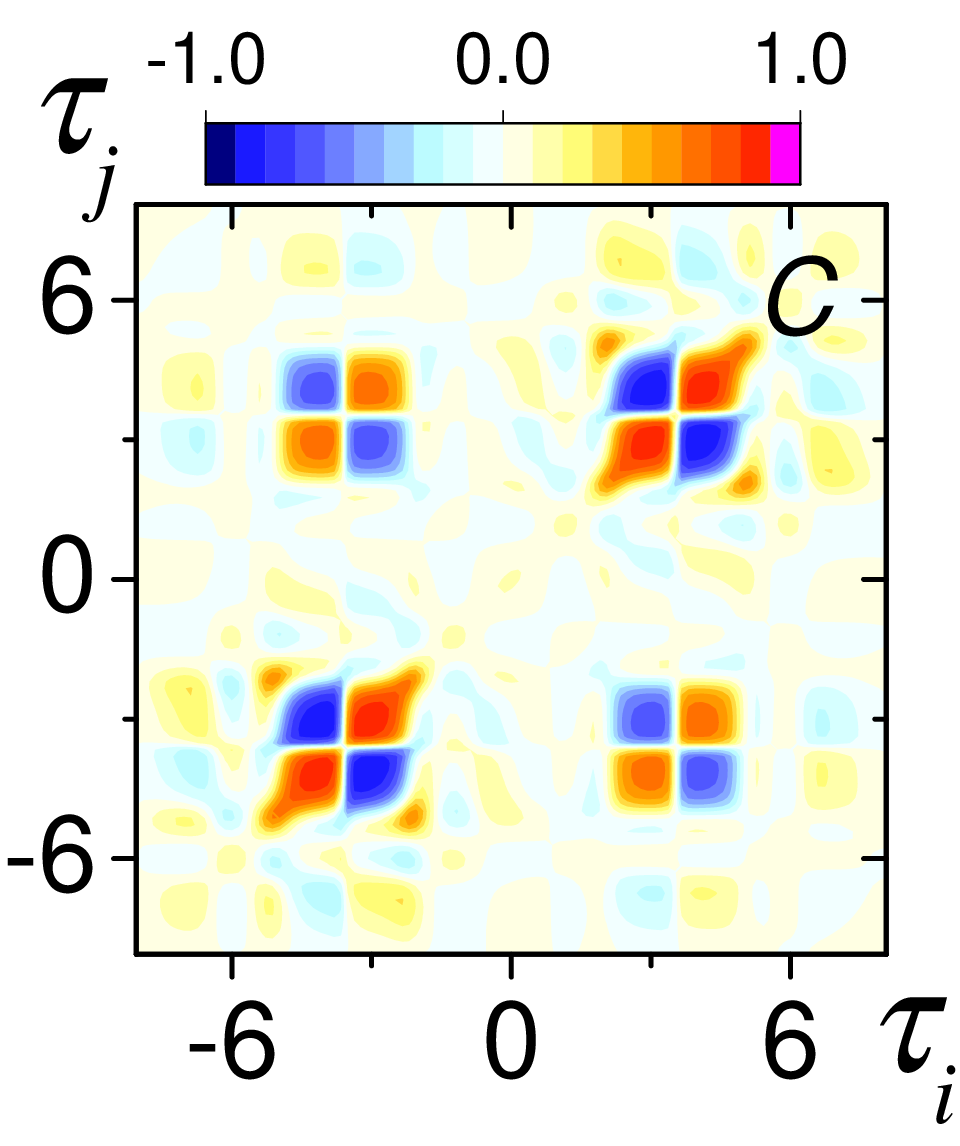} \\
(e) & (f) & (g) & (h)\\
\includegraphics[width=0.8in]{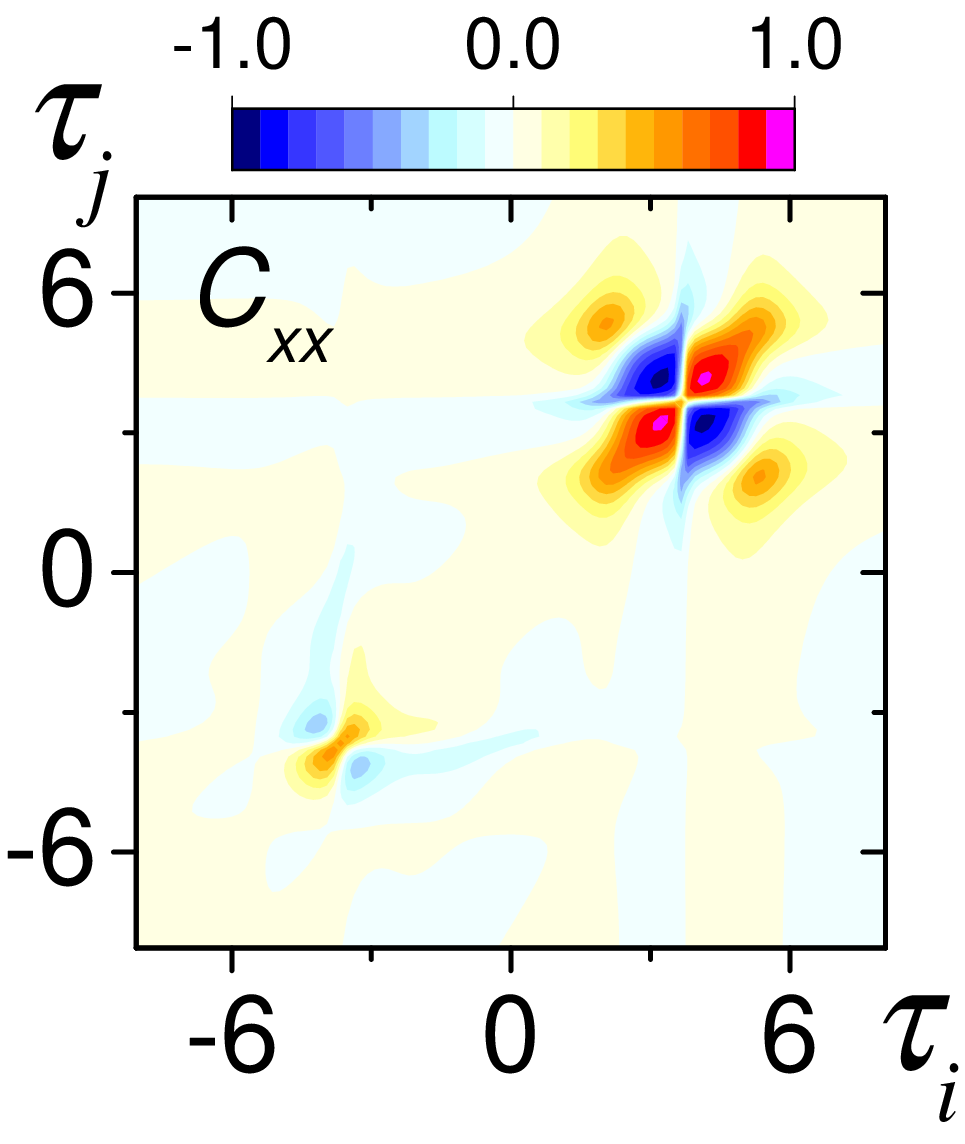} & 
\includegraphics[width=0.8in]{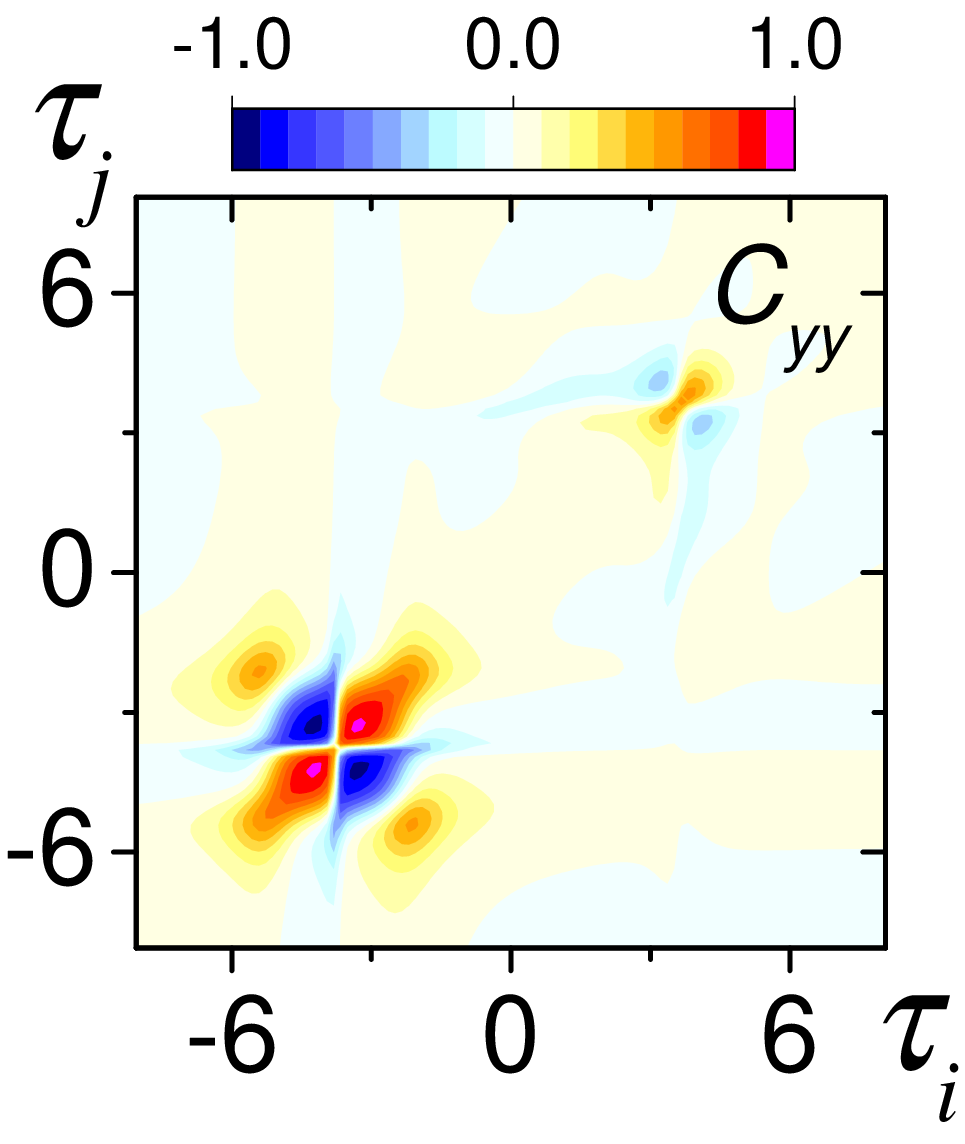} &
\includegraphics[width=0.8in]{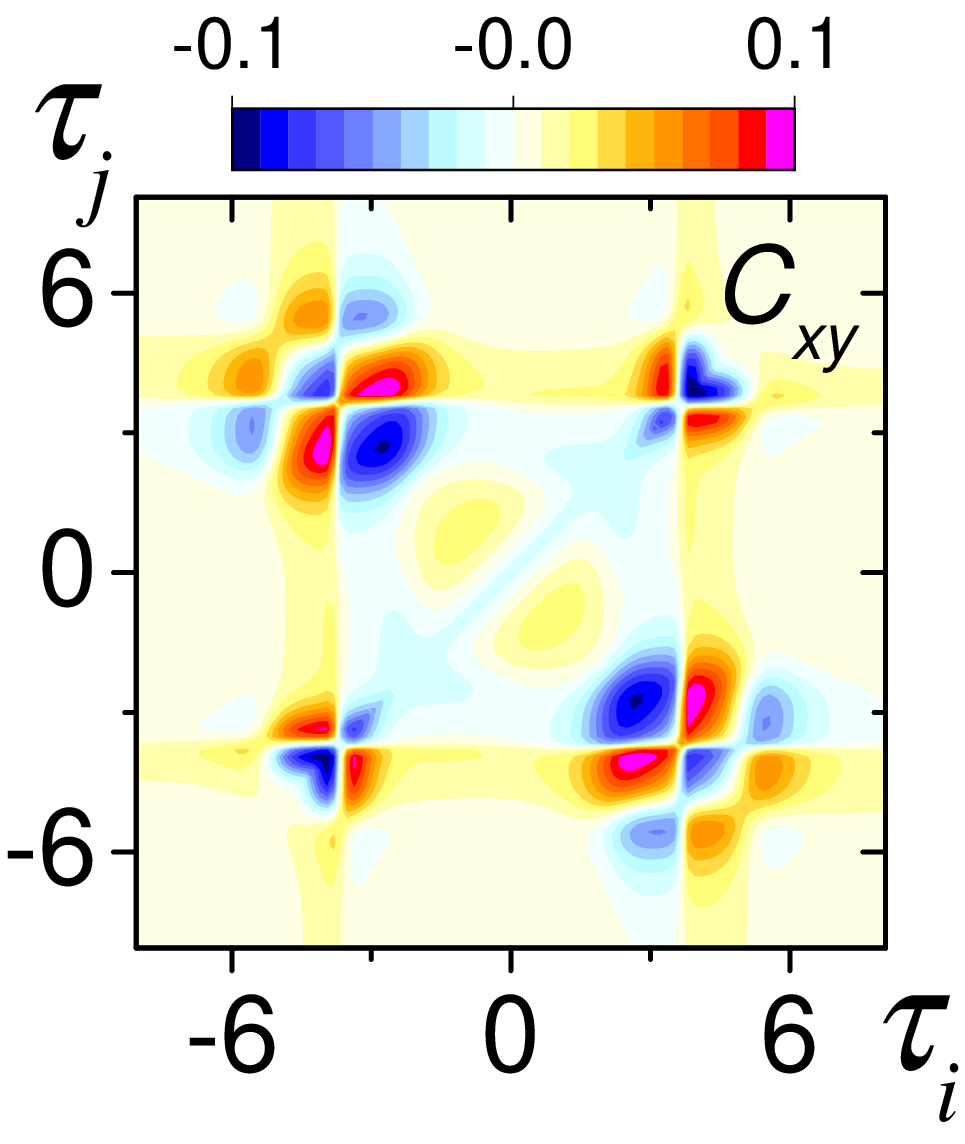} & 
\includegraphics[width=0.8in]{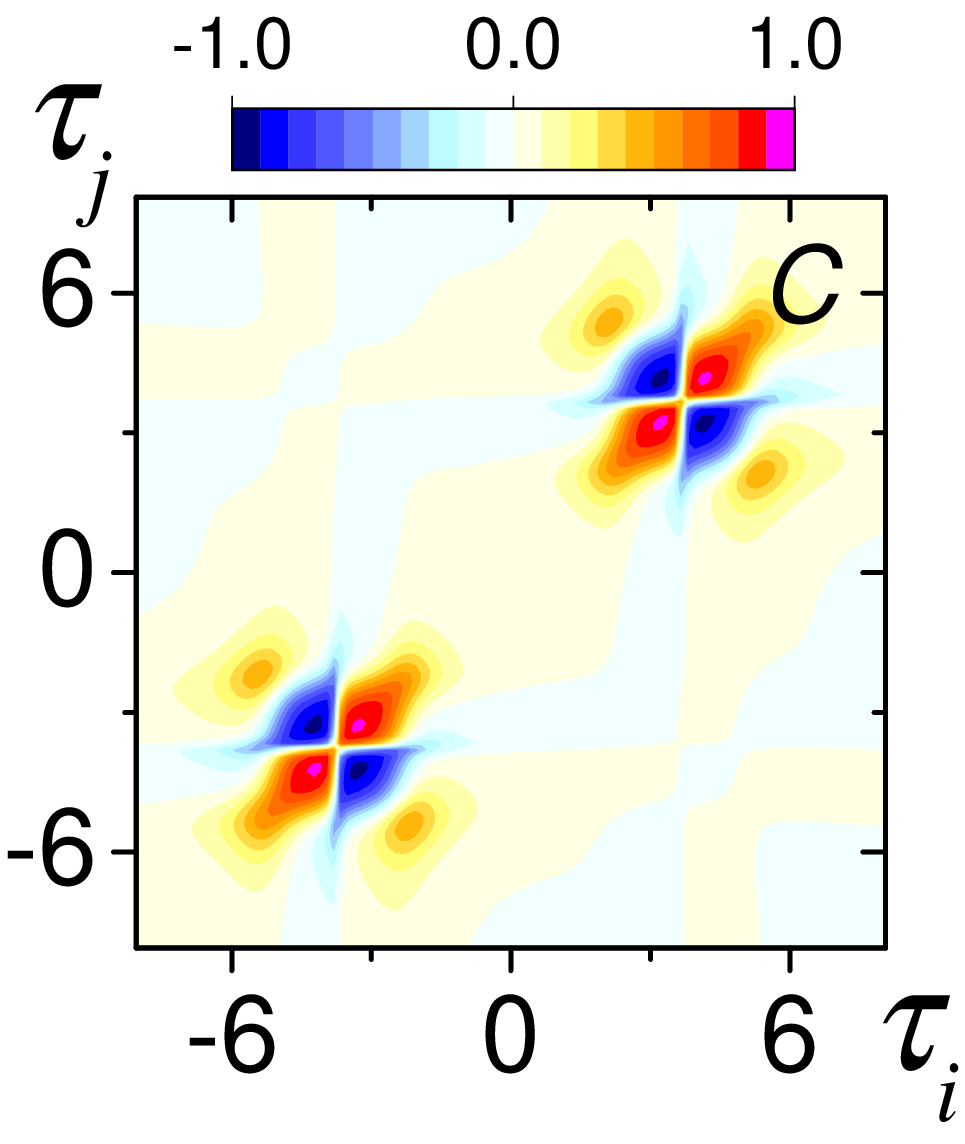}
\end{tabular}
\caption{ \label{fig5}
Time-domain correlation functions. 
(a),(b),(c), and (d) are calculated for sine-wave modulation (\ref{sinetr}) at the modulation period $\zeta_m=1.3$.
(e),(f),(g), and (h) are calculated for a fiber without modulation, $\zeta_m=\infty$.
The correlation functions $C_{kn}(\tau_i,\tau_j)$, $k,n=x,y$, and the complete correlation function $C(\tau_i,\tau_j)$ are given by Eq.~(\ref{Correlxxyy}) and Eq.~(\ref{Correl}), respectively. Other parameters are the same as in Fig.~\ref{fig4}. }
\end{figure}

Figure \ref{fig4} shows the collision of two frequency-shifted pulses. The dispersion modulation is given by Eq. (\ref{sine}). The pulses propagate with different group velocities and collide at a propagation distance $\zeta=3$. The group velocities of the pulses and their polarization are maintained after the collision. However, the correlations between pulses change greatly. 
The function $C_{xx}$ shows strong intrapulse correlations for $x$-polarized pulse [Fig.~\ref{fig5}(a)], and $C_{yy}$ for $y$-polarized one [Fig.~\ref{fig5}(b)]. 
The interpulse correlation appears in the distribution of the function $C_{xy}$  [Fig.~\ref{fig5}(c)]. The complete correlation function $C$ shows both intrapulse and interpulse correlations [Fig.~\ref{fig5}(c)]. The function $C$ achieves peak values $\pm 0.95$ for intrapulse correlations (diagonal patterns) and $\pm 0.7$ for interpulse correlations (off-diagonal patterns).    

To study the effect of dispersion modulation on photon-number correlations, I calculate the propagation of pulses in a fiber without dispersion modulation. Figure \ref{fig4} shows pulse collision in the presence of modulation. Without modulation, the pulses collide similarly, and the correlation functions $C_{xx}$ and $C_{yy}$ qualitatively remain the same. For comparison, see Figs.~\ref{fig5}(a) and \ref{fig5}(e) for $C_{xx}$ and Figs.~\ref{fig5}(b) and \ref{fig5}(f) for $C_{yy}$.
The modulation significantly affects the function $C_{xy}$. In the presence of modulation, the function $C_{xy}$ reaches peak values $\pm 0.7$ [Fig.~\ref{fig5}(c)]. Without modulation, the peak values of $C_{xy}$ reduce to $\pm 0.1$.

The effect of dispersion modulation on photon-number correlations can be clearly observed by comparing the complete correlation functions $C(\tau_i,\tau_j)$ calculated for a fiber with and without dispersion modulation. Figures~\ref{fig5}(d) and \ref{fig5}(h) show that dispersion modulation induces interpulse correlation. Correlations arise after pulse collision. I found that dispersion modulation produces persistent interpulse correlations when the spectra of the initial pulses overlap. With non-overlapping pulse spectra, correlations arise only at the collision point and disappear with further propagation of the pulses, as reported for scalar solitons in \cite{Konig2002}.

\section{Birefringent fiber \label{sec4}}

In this section, I consider the pulse propagation governed by two coupled NLS Eqs.~(\ref{schroedinger}) with the coefficients $A=1$, $B=2/3$, $C=1/3$, nonzero differential group delay $b_1$, and birefringence $b$. A steady-state vector soliton solition of  Eqs.~(\ref{schroedinger}) can be found only under specific conditions \cite{Agrawal2013, akhmediev1997}. However, the pulses exhibit soliton-like behavior, e.g., they tend to preserve their shape when the pulse energy exceeds a specific value. 

\begin{figure}
\begin{tabular}{ll}
(a) & (b)\\
\includegraphics[height=1.1in]{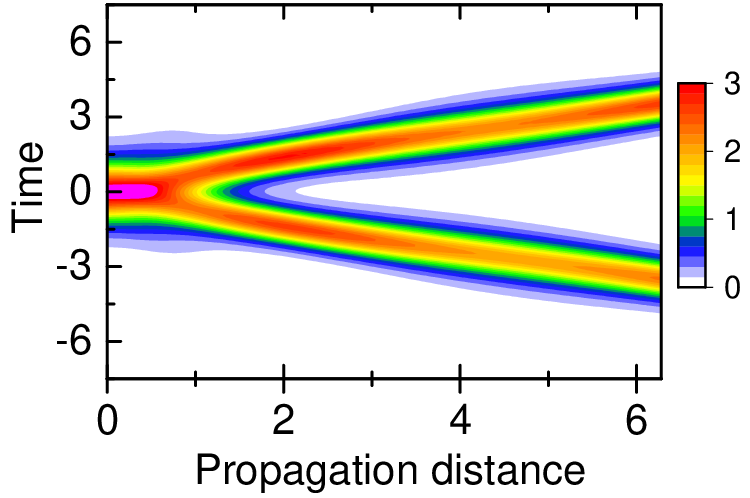} &
\includegraphics[height=1.1in]{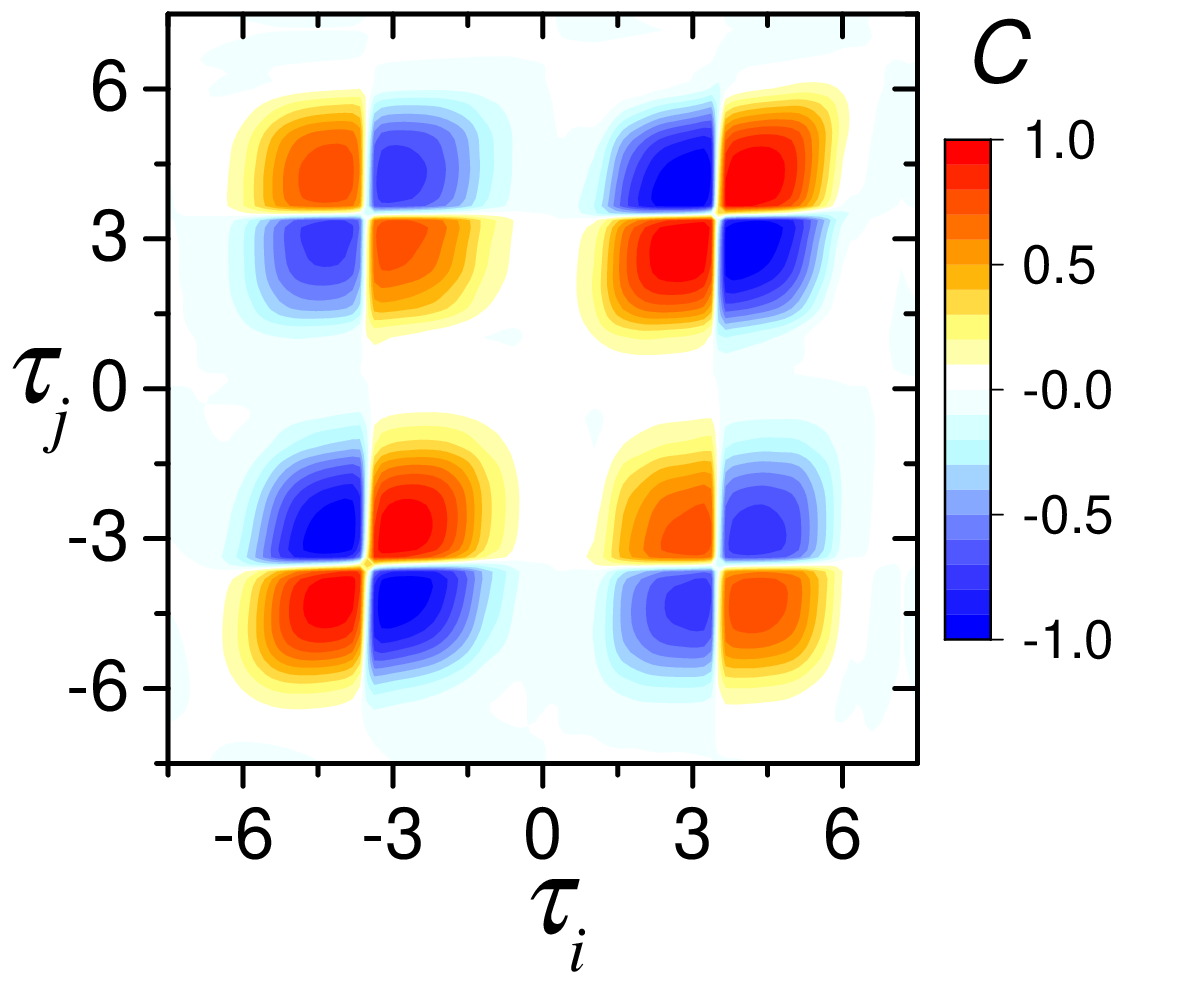} \\
(c) & (d)\\
\includegraphics[height=1.1in]{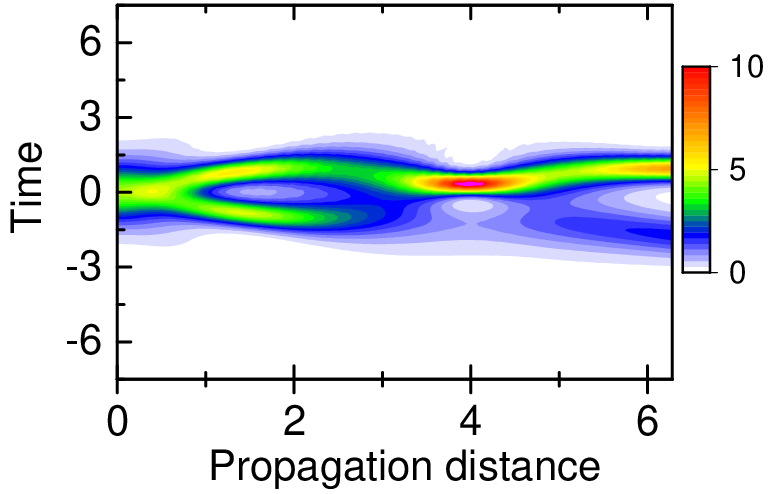} & 
\includegraphics[height=1.1in]{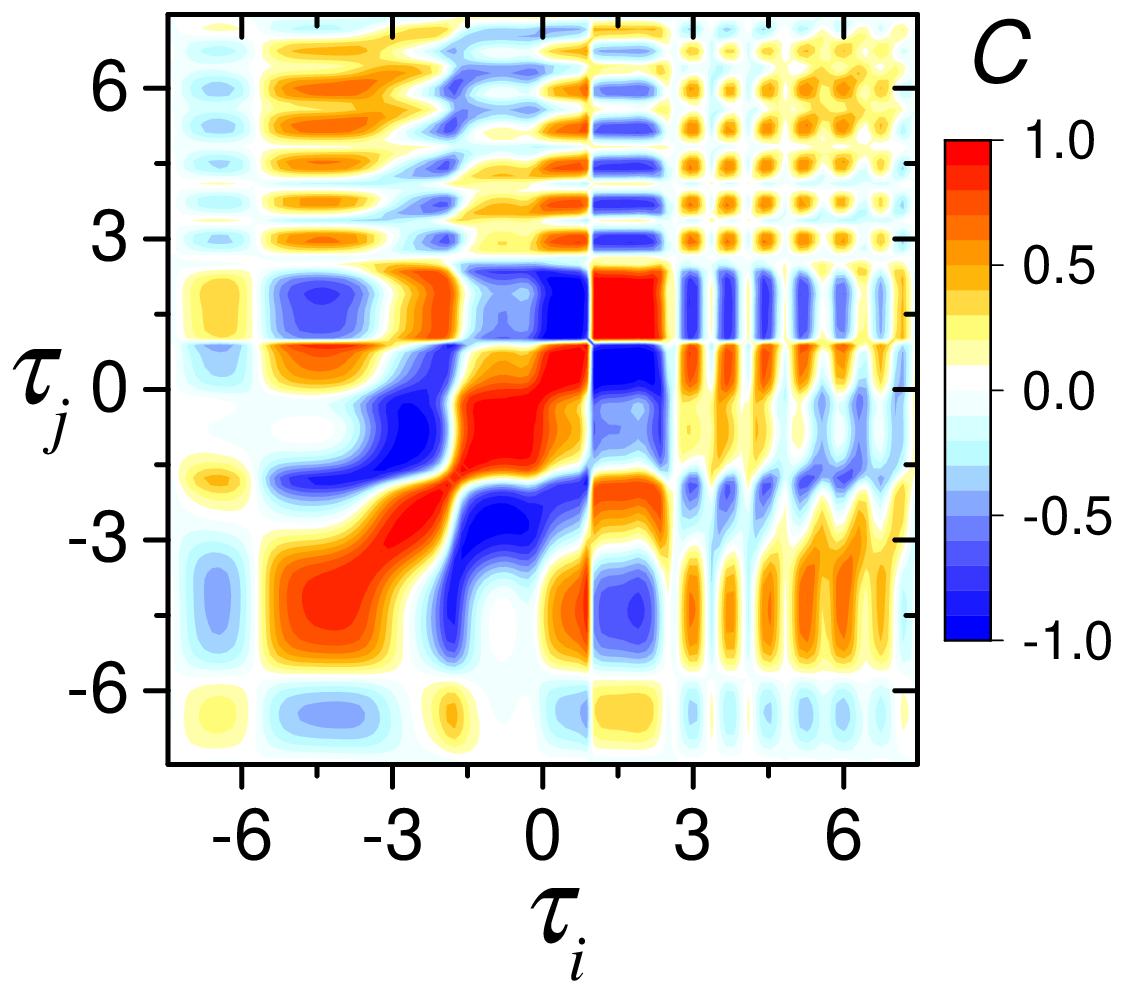} \\
(e) & (f)\\
\includegraphics[height=1.1in]{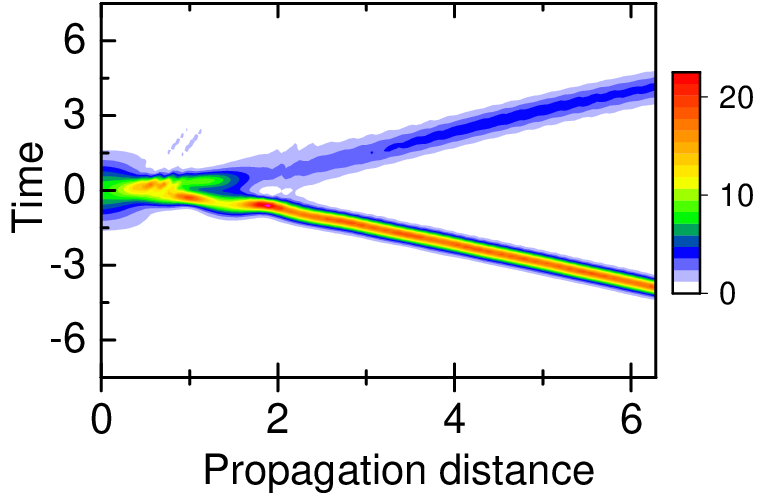} & 
\includegraphics[height=1.1in]{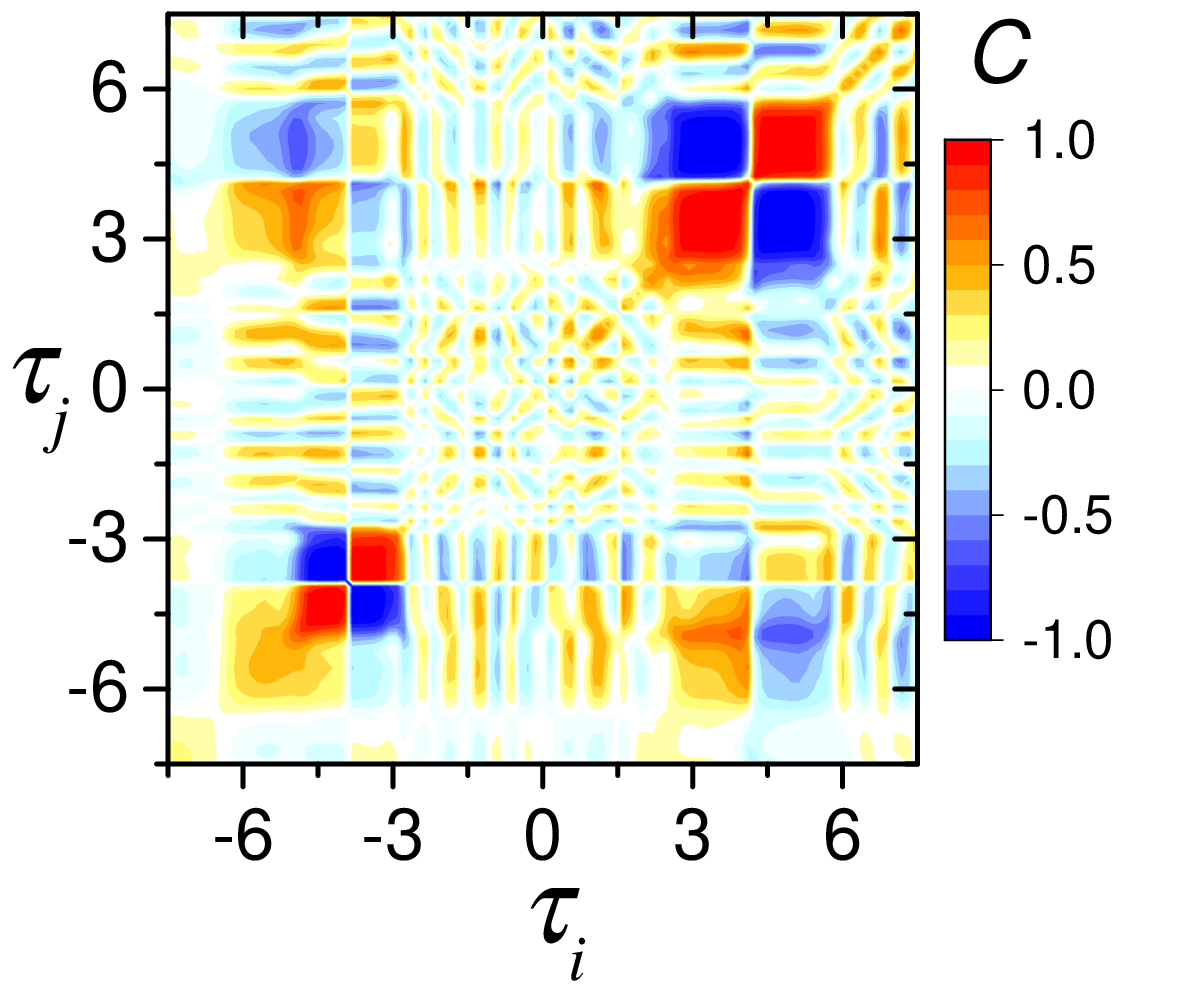} \\
\end{tabular}
\caption{
\label{fig6}
Pulse splitting due to PMD. 
(a),(c), and (e) False color plots of the field intensity $I(\zeta,\tau)=|U_x|^2+|U_y|^2$. 
(b),(d), and (f) The complete correlation function $C(\tau_i,\tau_j)$ calculated at a distance $\zeta=L=2\pi$. (a) and (b)  $u_0=1.8$.  (c) and (d)  $u_0=2.12$.  (e) and (f)  $u_0=2.83$. 
Other parameters are $\zeta_m=\infty$, $b_1=2$, and $b=20$. 
}
\end{figure}

Figure \ref{fig6} shows the pulse splitting under the effect of PMD. The initial field for the numerical simulations is given by Eq. (\ref{Uini}). At $u_0=1.8$, the PMD splits the pulse into fast and slow polarization components propagating at different group velocities [Fig.~\ref{fig6}(a)]. The pulses have sufficient energy to maintain their shape due to soliton-like  propagation. One of the pulses is $x$-polarized, and the other is $y$-polarized. These pulses can be processed separately using polarization beam splitter. 
The complete correlation function shows both high intrapulse  and interpulse correlations [Fig.~\ref{fig6}(b)]. The function $C(\tau_i,\tau_j)$ achieves peak values $\pm 0.96$ for intrapulse correlations (diagonal patterns) and $\pm 0.73$ for interpulse correlations (off-diagonal patterns). 

With increasing input power, the process of pulse propagation changes drastically due to polarization instability \cite{Agrawal2013}. Nonlinear birefringence reduces intrinsic birefringence. As a result, the pulse splitting due to PMD becomes ineffective [Fig.~\ref{fig6}(c)]. The structure of the complete correlation function becomes complex as it is defined by two overlapping pulses [Fig.~\ref{fig6}(d)].

With a further increase in the input power, nonlinear birefringence becomes dominant. After splitting, we obtain high-intensity slow pulse and low-intensity fast one [Fig.~\ref{fig6}(e)]. The asymmetry in the pulse splitting arises due to the energy transfer from the $y$-component of the pulse to the $x$-component. The interpulse correlation reaches the peak value of $C=0.7$ about $(\tau_i,\tau_j)=(4,-4.8)$ and $C=-0.7$ about $(\tau_i,\tau_j)=(5,-4.8)$ [Fig.~\ref{fig6}(f)]. The interpulse correlations are nonzero in narrow temporal regions. The intrapulse correlations have values $|C|>0.9$ in broad regions located around points $(\tau_i,\tau_j)=(-3.8,-3.8)$ and $(\tau_i,\tau_j)=(4.1,4.1)$. These points correspond to the maximum intensity of the output pulses.

\begin{figure}
\begin{tabular}{ll}
(a) & (b)\\
\includegraphics[height=1.1in]{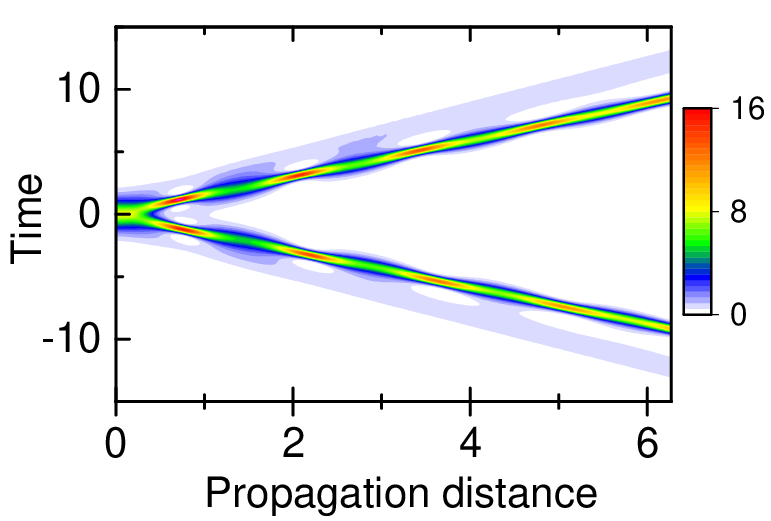} &
\includegraphics[height=1.1in]{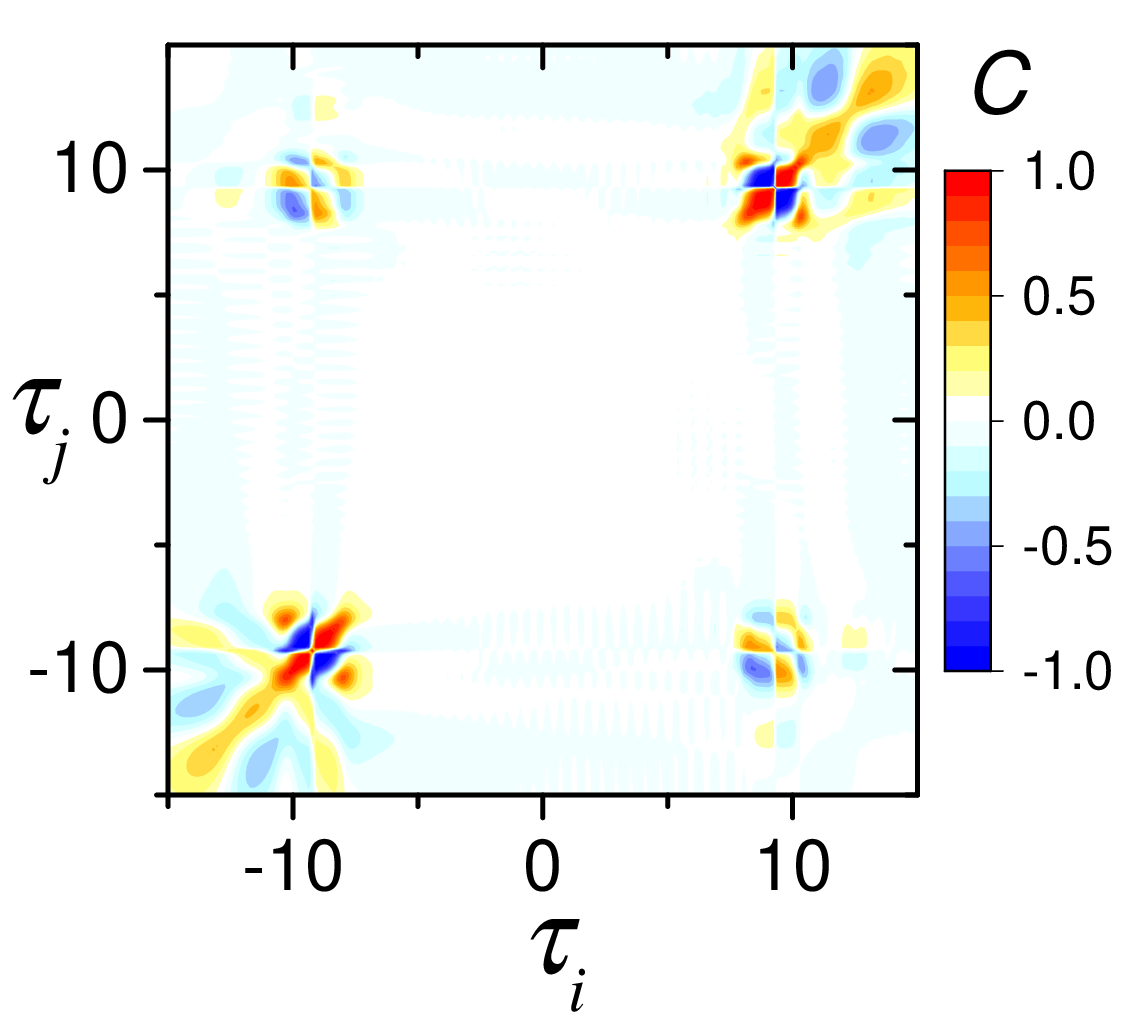} \\
(c) & (d)\\
\includegraphics[height=1.1in]{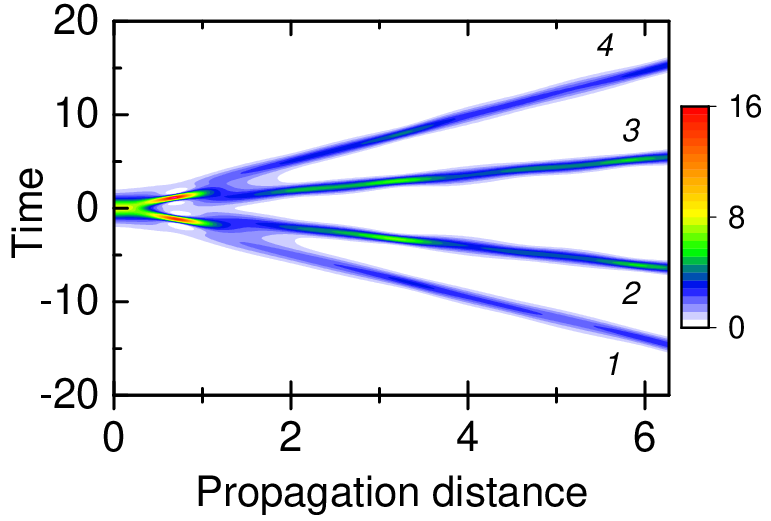} & 
\includegraphics[height=1.1in]{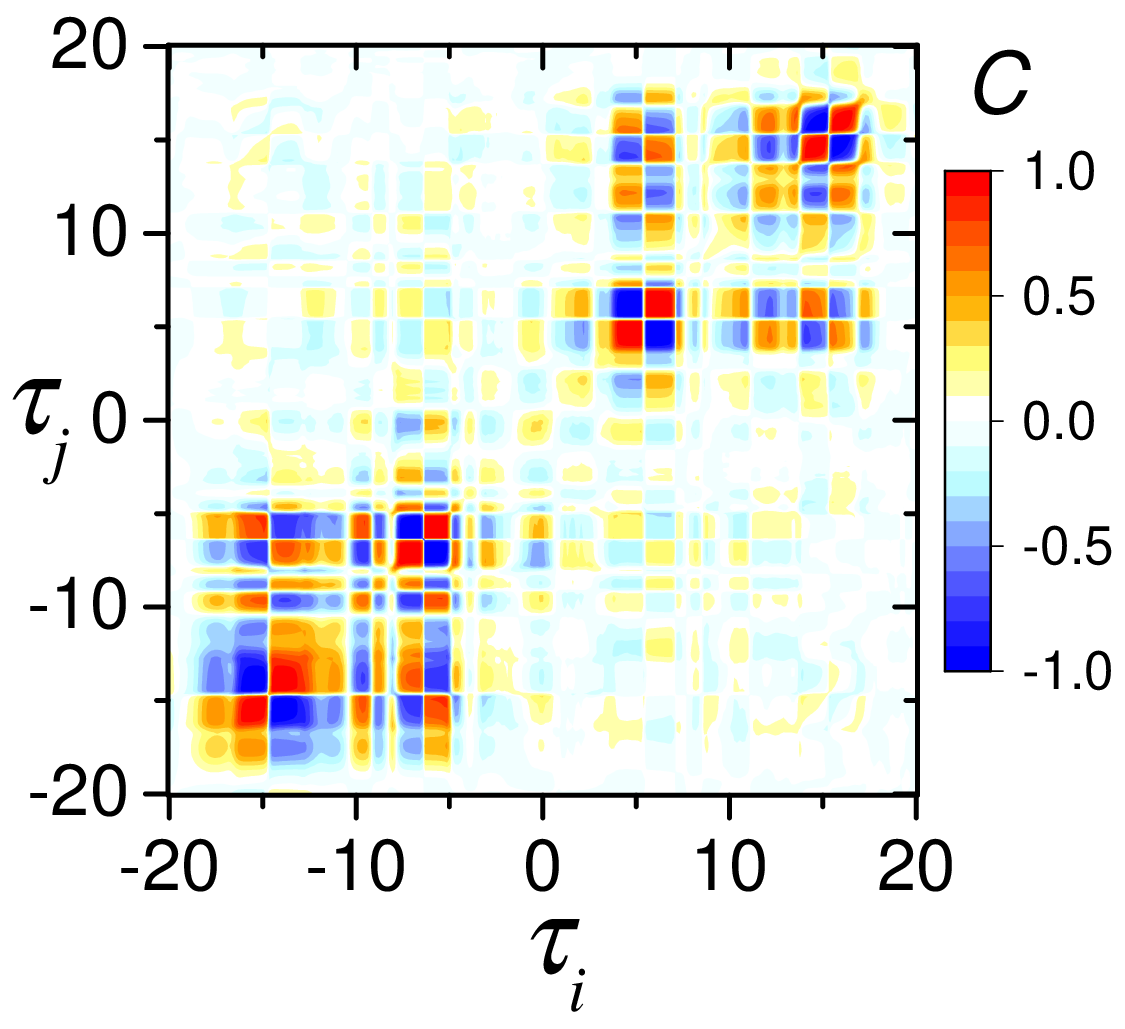}
\end{tabular}
\caption{
\label{fig7}
Pulse splitting due to PMD and dispersion modulation. 
(a) and (c) False color plots of the field intensity $I(\zeta,\tau)=|U_x|^2+|U_y|^2$. 
(b) and (d) The complete correlation function $C(\tau_i,\tau_j)$ calculated at a distance $\zeta=L=2\pi$. 
(a) and (b)  $\zeta_m=\infty$.  (c) and (d)  $\zeta_m=1.3$. 
Other parameters are $u_0=2.83$, $b_1=4$, and $b=40$. 
}
\end{figure}

The destructive effect of polarization instability on the pulse splitting can be reduced with high values of PMD. Figure \ref{fig6} is calculated at $b_1=2$ and $b=20$. At $b_1=4$ and $b=40$, the initial pulse is split into two identical pulses in all three regimes ($u_0=1.8$, $u_0=2.12$, and $u_0=2.83$) shown in Fig.~\ref{fig6}. Fig.~\ref{fig7}(a) shows pulse splitting at $u_0=2.83$ and $b_1=4$. A high PMD induces symmetric splitting of the pulse, i.e., slow and fast output pulses have the same amplitude. However, increasing $b_1$ reduced the interpulse correlations [Fig.~\ref{fig7}(b)].

After splitting, the energy of each pulse is sufficient to support a breather-like pulse propagation [Fig.~\ref{fig7}(a)]. The amplitude and width of each pulse vary quasi-periodically as it propagates, and the pulses maintain their shape. Modulation of the fiber dispersion can cause the splitting of the breather. Fig.~\ref{fig7}(c) shows the pulse propagation under the simultaneous effect of the PMD and dispersion modulation. I consider the weak modulation of the group delay and fiber birefringence
\begin{eqnarray} \label{deltab}
b_1(\zeta)&=&4(1+0.01\sin(2\pi \zeta/\zeta_m)),\\
b(\zeta)&=&40(1+0.01\sin(2\pi \zeta/\zeta_m)). 
\end{eqnarray}
The normalized group velocity dispersion parameter $D(\zeta)$ is given by Eq.~(\ref{sine}).  

With dispersion modulation, pulse splitting occurs in two stages [Fig.~\ref{fig7}(c)]. In the first stage, $\zeta<0.4$, the pulse is split into two polarization components due to PMD. In the second stage, $\zeta =1.0$, each polarization component is split into two pulses due to modulation of the dispersion parameter $D(\zeta)$. At the output we have four pulses. Pulses \textsl{1} and \textsl{2} are $x$-polarized, and pulses \textsl{3} and \textsl{4} are $y$-polarized [Fig.~\ref{fig7}(c)]. At $\zeta>2$ the pulses propagate as fundamental solitons that cannot be split by dispersion modulation. For fundamental solitons, the dispersion modulation leads to the generation of low-intensity dispersive waves, which create fringes in the correlation pattern [Fig.~\ref{fig7}(d)]. 

Correlations arise between the pulses with the same polarization states, namely, between the pulses \textsl{1} and \textsl{2} $(-20\le\tau_{i,j}\le0)$, and between pulses \textsl{3} and \textsl{4} $(0<\tau_{i,j}\le 20)$ [Fig.~\ref{fig7}(d)]. There are no correlations between pulses \textsl{1} and \textsl{3} or between pulses \textsl{2} and \textsl{4}. At large values of the differential group delay $b_1$, the spectra of pulses with orthogonal polarization states become non-overlapping. As a result, the correlations between these pulses disappear. This effect occurs during both pulse splitting and pulse collisions.

\section{Conclusion \label{sec5}}

In this paper, I considered the quantum correlations arising in the process of splitting of second-order solitons and inelastic soliton collisions stimulated by periodically modulated fiber dispersion. The backpropagation method \cite{haus1990, lai1993, lai1995, Rand2005} was applied to the study of quantum correlations and squeezing in vector soliton-like pulses. Two approaches are used in the modeling of the pulse propagation. The first approach is based on the Manakov equations. The second approach uses the coupled NLS equations, which include terms describing differential group delay and linear birefringence. 

In the Manakov equation model the splitting of a soliton can be obtained with dispersion modulation. The second-order vector soliton is split into two similar pulses, which have opposite frequency shift. Due to the frequency shift, the pulses propagate with different group velocities. 

In a fiber with dispersion modulation, steady-state vector solitons do not exist. However, each pulse exhibits soliton-like behavior, i.e., the pulses tend to preserve their shape and acquire a nonzero squeezing ratio. Soliton splitting produces pulses with nonzero intrapulse and interpulse correlations. The correlated pulses can be generated over a short propagation distance comparable to the soliton period.

Both the time-domain and frequency-domain correlations were calculated. In the time domain, the pulses produce a four-lobe correlation pattern similar to that of scalar solitons \cite{lee2005}. In the frequency domain, the correlation pattern has a checkerboard structure. This structure is connected to multiple local maxima that arise from the interference of the overlapping pulse spectra. The interpulse correlation is sensitive to the modulation period of the fiber dispersion. This opens a facility to control the correlation between pulses.

In the collision of two co-propagating pulses, the dispersion modulation changes the group velocities and polarization states of the output pulses. The pulses are polarized elliptically and have highly correlated photon-number noise. To test the effect of dispersion modulation on the interpulse correlations, I considered the collision of two frequency-shifted pulses propagating at different group velocities. The changes in the pulse shapes and spectra due to dispersion modulation are not significant. However, the interpulse correlations changed significantly. In the considered example, the interpulse correlation function reaches peak values $\pm0.7$ in the presence of dispersion modulation and $\pm0.1$ in the absence of modulation. When two unperturbed solitons collide, the interpulse correlations remain at the collision point and disappear as the solitons propagate \cite{Konig2002}. The dispersion modulation maintains the interpulse correlation after collision.

In the model of coupled NLS equations, the splitting of soliton-like pulses can be induced not only by dispersion modulation but also by PMD. PMD produces two linearly polarized pulses with highly correlated quantum noise. The pulses have an orthogonal polarization state and can be separated into two channels using a simple polarization beam splitter. The generation of highly correlated distinct pulses is possible at a certain relation between the initial pulse amplitude and differential group delay. High values of the differential group delay erase the interpulse correlations.

As a specific example, I studied the evolution of an optical pulse polarized linearly at $45^\circ$ from the $x$ axis of the coordinate system. Additionally, linearly and elliptically polarized pulses demonstrate similar behavior in pulse splitting and inelastic pulse collision.

\appendix
\section{Linearized propagation equations \label{app1}}

In this appendix, the explicit expressions for the evolution of the fluctuation operators are given. The operators $\hat{u}_x$ and $\hat{u}_y$ obey two coupled equations 

\noindent
\begin{equation}
\begin{array}{rl}
\displaystyle
  \frac{\partial \hat{u}_x}{\partial \zeta} = & -
  \displaystyle
  b_1(\zeta) \frac{\partial \hat{u}_x}{\partial \tau} + i
  \displaystyle
  \frac{D(\zeta)}{2} \frac{\partial^2 \hat{u}_x}{\partial \tau^2} + i b(\zeta) \hat{u}_x 
\\*[\medskipamount]
  &+i (2A |U_{x}|^2 + B |U_{y}|^2) \hat{u}_x
\\*[\medskipamount]
  &+i (A U_{x}^2 + C U_{y}^2) \hat{u}_x^\dagger
\\*[\medskipamount]
  &+i (2C U_{x}^\ast U_{y} + B U_{x} U_{y}^\ast) \hat{u}_y
\\*[\medskipamount]
  &+i B U_{x} U_{y} \hat{u}_y^\dagger,
\end{array}
\label{u1}
\end{equation}
\noindent

\begin{equation}
\begin{array}{rl}
\displaystyle
  \frac{\partial \hat{u}_y}{\partial \zeta} = & 
  \displaystyle 
  b_1(\zeta) \frac{\partial \hat{u}_y}{\partial \tau} + i
  \displaystyle
  \frac{D(\zeta)}{2} \frac{\partial^2 \hat{u}_y}{\partial \tau^2} 
- i b(\zeta) \hat{u}_y 
\\*[\medskipamount]
  &+i (2A |U_{y}|^2 + B |U_{x}|^2) \hat{u}_y
\\*[\medskipamount]
  &+i (A U_{y}^2 + C U_{x}^2) \hat{u}_y^\dagger
\\*[\medskipamount]
  &+i (2C U_{y}^\ast U_{x} + B U_{y} U_{x}^\ast) \hat{u}_x
\\*[\medskipamount]
  &+i B U_{x} U_{y} \hat{u}_x^\dagger.
\end{array}
\label{u2}
\end{equation}

\vspace{1.5cm}
The adjoint system is defined by following equations: 

\noindent
\begin{equation}
\begin{array}{rl}
\displaystyle
  \frac{\partial u^A_x}{\partial \zeta} = & - \Bigl( 
  \displaystyle
  b_1(\zeta) \frac{\partial u^A_x}{\partial \tau} 
 -i\frac{D(\zeta)}{2} \frac{\partial^2 u^A_x}{\partial \tau^2} 
 -i b(\zeta) u^A_x 
\\*[\medskipamount]
  &-i (2A |U_{x}|^2 + B |U_{y}|^2) u^A_x
\\*[\medskipamount]
  &+i (A U_{x}^2 + C U_{y}^2) u^{A\ast}_x
\\*[\medskipamount]
  &-i (2C U_{x}^\ast U_{y} + B U_{x} U_{y}^\ast) u^A_y
\\*[\medskipamount]
  &+i B U_{x} U_{y} u^{A\ast}_y \Bigl),
\end{array}
\label{uA1}
\end{equation}

\noindent
\begin{equation}
\begin{array}{rl}
\displaystyle
  \frac{\partial u^A_y}{\partial \zeta} = & - \Bigl( 
  \displaystyle
 -b_1(\zeta) \frac{\partial u^A_y}{\partial \tau}
 -i\frac{D(\zeta)}{2} \frac{\partial^2 u^A_y}{\partial \tau^2} 
 +i b(\zeta) u^A_y 
\\*[\medskipamount]
  &-i (2A |U_{y}|^2 + B |U_{x}|^2) u^A_y
\\*[\medskipamount]
  &+i (A U_{y}^2 + C U_{x}^2) u^{A\ast}_y
\\*[\medskipamount]
  &-i (2C U_{y}^\ast U_{x} + B U_{y} U_{x}^\ast) u^A_x
\\*[\medskipamount]
  &+i B U_{x} U_{y} u^{A\ast}_x \Bigl).
\end{array}
\label{uA2}
\end{equation}

\phantom{AAA}

\section{Homodyne detection scheme \label{app2}}

This appendix describes a method for detecting quantum fluctuations of polarized optical pulses, more specifically, a method for detecting the measurement operator $\hat{M}$ introduced in Section \ref{sec2}.

\begin{figure}[hb]
\includegraphics[width=2.2in]{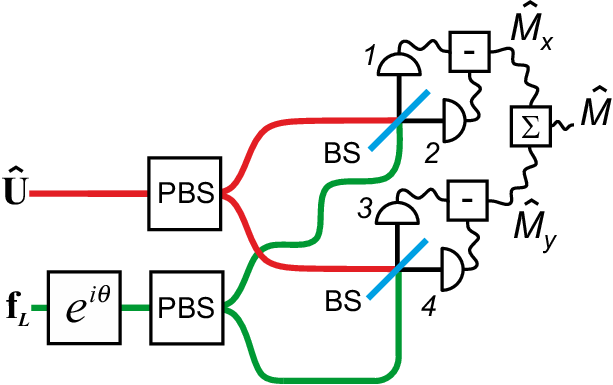}
\caption{\label{fig8} 
Schematic of the homodyne receiver employing polarization diversity.
${\bf \hat{U}}$ is the incoming signal, 
and ${\bf f}_{L}$ is the field of the LO.
The box labeled with $e^{i\theta}$ shows the phase adjustment for the LO field. 
PBS is a polarization beam splitter. BS is a 50-50 beam splitter.
}
\end{figure}

Figure \ref{fig8} shows the optical scheme used in coherent optical communications \cite{Kikuchi2010} and in quantum homodyne detection based on the polarization diversity technique \cite{Yuan2008, Rand2005}.
The phase of the LO field is adjusted via \fbox{$e^{i\theta}$} element (Fig.~\ref{fig8}) in order to achieve the best squeezing ratio. For the simplicity, the phase delay $\theta$ is taken the same for both $x$ and $y$ components of the LO field. The incoming signal ${\bf \hat{U}}$ and LO field are split on a polarizing beam splitters. Each of the two outputs is detected by a pair of photodiodes. The difference in photocurrents of two photodiodes, e.g., \textsl{1} and \textsl{2} (Fig.~\ref{fig8}), yields the value for the operator \cite{haus1990}
\begin{equation} \label{Mx} 
\hat{M}_x(\zeta) =  \frac{1}{2}      
\int f_{Lx}^\ast(\tau) \hat{u}_x(\zeta,\tau) d\tau +\text{h.c.},
\end{equation}
where $f_{Lx}$ is the $x$-component of the electric field vector of the LO. A similar expression can be written for the operator $\hat{M}_y$ operator, which is determined by the difference in the photocurrents of photodiodes \textsl{3} and \textsl{4} (Fig.~\ref{fig8}).


The measurement operator $\hat{M}$ is
\begin{equation} \label{Mglobal}
\hat{M}(\zeta) = \hat{M}_x + \hat{M}_y =  
\langle {\bf f}_L | {\bf \hat{u}} \rangle,
\end{equation}
where ${\bf f}_L=(f_{Lx},f_{Lx}^\ast,f_{Ly},f_{Ly}^\ast)^T$ 
is the field of the LO, ${\bf \hat{u}}$ is given by (\ref{u}), and 
$\langle {\bf f}_L | {\bf \hat{u}} \rangle$ is the inner product.

The element \fbox{$\Sigma$} in Fig.~\ref{fig8} is introduced only to demonstrate the physical meaning of the measurement operator $\hat{M}$. 

\begin{acknowledgments}
The author is grateful to Dr. Ju. Ko\-nyukho\-va for the help in preparation of this article
and to Prof. L. Melnikov for useful discussions. This work has been supported by the Russian Science Foundation (grant No. 22-12-00396) https://rscf.ru/project/22-12-00396/
\end{acknowledgments}


\providecommand{\noopsort}[1]{}\providecommand{\singleletter}[1]{#1}%

\end{document}